\documentclass[9pt,twocolumn,twoside]{pnas-new}

\setboolean{displaywatermark}{false}
\usepackage{multirow}
\usepackage{setspace}

\def \T{\mathbf{T}}
\def \x{\mathbf{x}}

\templatetype{pnasresearcharticle} 

\title{Fast flow-based algorithm for creating density-equalizing map projections}

\author[a,1]{Michael~T.~Gastner}
\author[b]{Vivien Seguy} 
\author[a]{Pratyush More}

\affil[a]{Yale-NUS College, Division of Science, 16 College Avenue West, \#01-220 Singapore 138527}
\affil[b]{Graduate School of Informatics, Kyoto University, 36-1 Yoshida-Honmachi, Sakyo-ku, Kyoto 606-8501 Japan}
\leadauthor{Gastner} 

\significancestatement{
Geographic maps are a popular means to visualize spatial statistics.
Conventionally, each map region is displayed with an area proportional to the actual land area.
But equal-area maps can grossly misrepresent demographic data: densely populated cities should be given more prominence than large, but sparsely populated territories.
Cartograms solve this problem by rescaling map regions in proportion to, for example, population or gross domestic products.
Until now, it has generally been cumbersome or slow to calculate map projections for contiguous cartograms.
Here we describe and benchmark a fast flow-based algorithm that computes cartograms in a matter of seconds, yet
maintains the strengths of previous methods -- a development which may
lead to a more widespread adoption of cartograms.
}

\authorcontributions{M.T.G. designed research and analyzed
  data. M.T.G., V.S. and P.M. performed research and wrote the paper.}
\authordeclaration{The authors declare no conflict of interest.}
\correspondingauthor{\textsuperscript{1}To whom correspondence should be addressed. E-mail: michael.gastner@yale-nus.edu.sg}


\begin{abstract}
Cartograms are maps that rescale geographic regions (e.g., countries,
districts) such that their areas are proportional to quantitative
demographic data (e.g., population size, gross domestic product).
Unlike conventional bar or pie charts, cartograms can represent
correctly which regions share common borders, resulting in insightful 
visualizations that can be the basis for further spatial statistical
analysis.
Computer programs can assist data scientists in preparing cartograms,
but developing an algorithm that can quickly transform every
coordinate on the map (including points that are not
exactly on a border) while generating recognizable images has
remained a challenge.
Methods that translate the cartographic deformations into
physics-inspired equations of motion have become popular, but
solving these equations with sufficient accuracy can
still take several minutes on current hardware.
Here we introduce a flow-based algorithm whose equations of motion are numerically easier to solve compared with previous methods. 
The equations allow straightforward parallelization so that the calculation takes only a few seconds even for complex and detailed input.
Despite the speedup, the proposed algorithm still keeps the advantages of previous techniques: with comparable quantitative measures of shape distortion, it accurately scales all areas, correctly fits the regions together and
generates a map projection for every point.
We demonstrate the use of our algorithm with applications to the 2016
US election results, the gross domestic products of Indian states and
Chinese provinces, and the spatial distribution of deaths in the London borough of Kensington and Chelsea
between 2011 and 2014.
\end{abstract}

\dates{This manuscript was compiled on \today}
\doi{\url{www.pnas.org/cgi/doi/10.1073/pnas.XXXXXXXXXX}}

\begin{document}

\verticaladjustment{-2pt}

\maketitle
\ifthenelse{\boolean{shortarticle}}{\ifthenelse{\boolean{singlecolumn}}{\abscontentformatted}{\abscontent}}{}

\begin{figure}
\centering
\includegraphics[width=\linewidth]{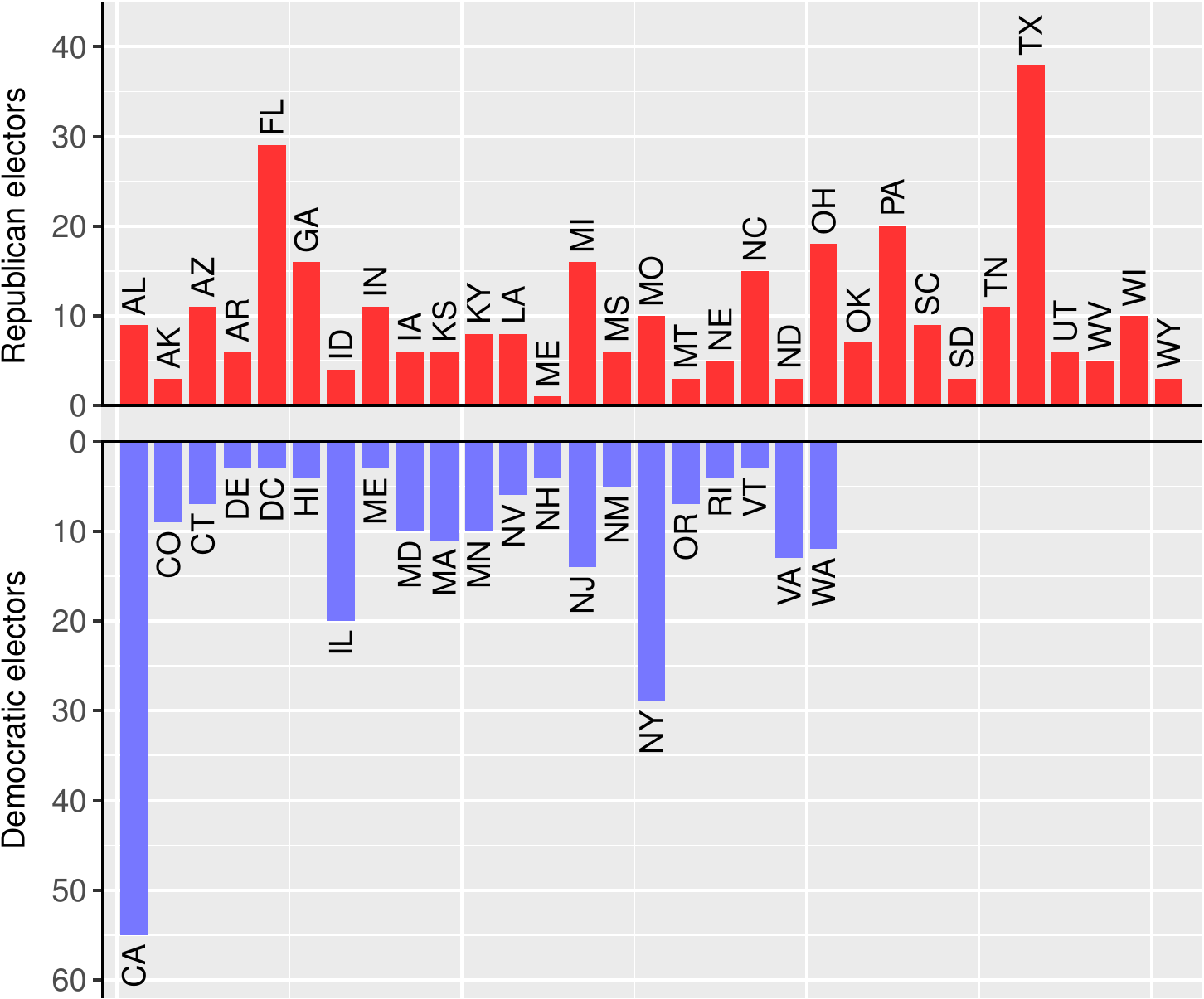}
\caption{
  A bar chart of the Electoral College vote for the US president
  in 2016. This diagram satisfies the area principle: the area of each
  bar is proportional to the number of electors.
  However, from this bar chart it is not clear where states are located geographically.
}
\label{barchart}
\end{figure}

\dropcap{A} guideline for displaying statistical data in a
diagram is the ``area principle'': each part of the diagram should have
an area in proportion to the number it
represents~\cite{Deveaux_etal16}.
For many categorical data, a bar chart is a simple visualization
method that satisfies the area principle. 
For example, if our data are the electors that voted for the US
president in December 2016, we can categorize the electors by US state.
Every bar in the top half of Fig.\,\ref{barchart} corresponds
to a state that sent at least one Republican elector to the Electoral College.
In the bottom half, the bars show the states with Democratic electors.
The colors chosen for the bars are the traditional red for Republicans
and blue for Democrats.
Because the bar chart satisfies the area principle, the election is
won by the color that occupies more area, which is evidently red in
this example.\footnote{Because of peculiarities in the US electoral
  system, the Electoral College is not an exact representation of the
  proportion of votes cast by the US population at large. The
  Republican candidate Donald Trump was elected as US president
  despite losing the popular vote to the Democratic candidate Hillary
  Clinton. We show a cartogram of the popular vote distribution in
  Fig.~\ref{uspopvote} of the SI Appendix.}

\begin{figure}
\centering
\includegraphics[width=\linewidth]{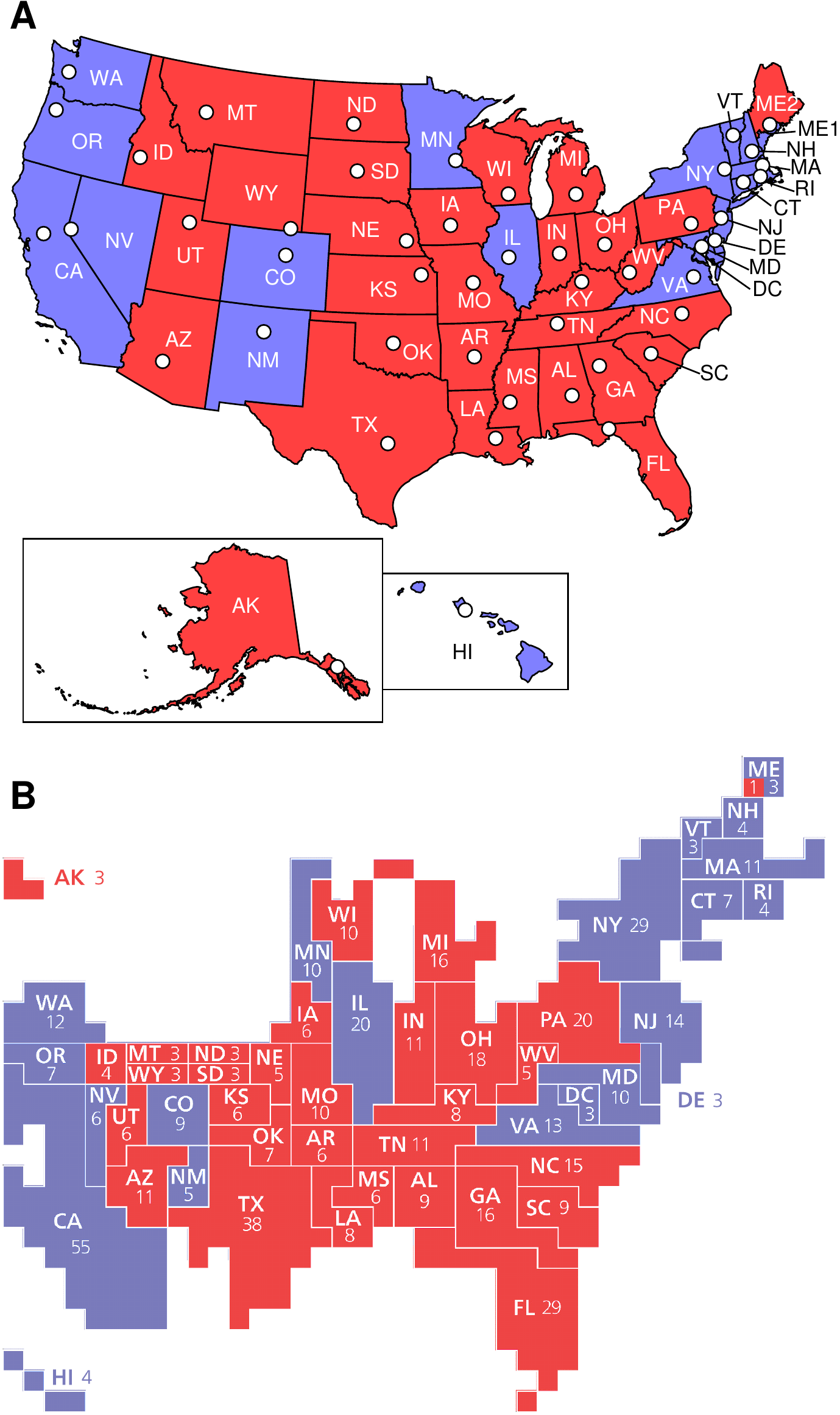}
\caption{
  (A) A conventional map projection (here an Albers projection)
  clearly shows the location of each state, but violates the area
  principle:
  states that occupy a large area do not necessarily have a large
  number of electors.
  (B) A cartogram of the 2016 Electoral College [adapted from
  Wikipedia~\cite{WikipediaCart}] satisfies the area principle.
  Each elector is represented by a small square at the approximate
  location of the elector's home state.
  Cartograms such as these are popular in the media, but are not map
  projections in a strict sense: there is no continuous mathematical function that transforms coordinates of longitude and latitude to coordinates on the cartogram. For example, in (B) it is not
  possible to identify the location of the state capitals (indicated
  by white circles in panel A).
}
\label{electoral_college}
\end{figure}

Although a bar chart is often a suitable visualization tool, it
cannot reveal the spatial pattern behind the data.
The bar chart of Fig.~\ref{barchart} lacks the information where the
states are located: neighboring bars do not necessarily correspond to
states that are geographic neighbors.
If we want to visualize how the states fit together in real space,
we need a different approach.
The common alternative is to show a map such as
Fig.\,\ref{electoral_college}A, where we use an Albers equal-area conic
projection for the contiguous United States to produce their familiar geographic
outline.
We add Alaska and Hawaii, suitably rescaled, below the map of the
contiguous United States.
Each state is filled with either red or blue depending on the party
affiliation of the electors.\footnote{The only exception is Maine which applies
the ``congressional district method'': although the majority in Maine
voted for the Democratic candidate Hillary Clinton, the Republican
candidate Donald Trump still gained one electoral vote for winning the
2nd congressional district (abbreviated as ME2 in Fig.\,\ref{electoral_college}A.)}

The map in Fig.\,\ref{electoral_college}A accurately shows the
relative area and position of each state.
However, it does not obey the area principle of statistics.
For example, Montana (abbreviated by MT in
Fig.\,\ref{electoral_college}A) covers more than 2000 times the area of
Washington DC, but both regions have the same number of electors.
On aggregate, Republican electors won 74\% of the US area in square
kilometers, but had only 57\% of the vote share in the Electoral
College.
So, Fig.\,\ref{electoral_college}A has the opposite problem of
Fig.\,\ref{barchart} where we satisfied the statistical area
principle, but conveyed no information about the states' locations.
One might suspect that showing the locations and simultaneously
satisfying the area principle are as impossible as squaring the circle.
Fortunately, however, there is a visualization method, known as a
cartogram, that can tackle this challenge~\cite{Tobler04,
  NusratKobourov16}.

After a brief review and classification of cartograms, we
introduce a technique that produces cartograms of a quality comparable to
the most popular technique currently in use: the diffusion
cartogram~\cite{GastnerNewman04}.
The technique proposed in this article solves a completely different set of equations 
so that the computation can finish within a fraction of the previously needed time.
We benchmark our algorithm with data from the USA, India, China, and the London borough of Kensington and Chelsea to
demonstrate that our method accurately satisfies the area principle
and generates visually pleasing cartograms.

\section*{Classification of cartogram methods}

In a cartogram, regions are deformed such that their areas
are equal to statistical data such as population, votes in an
election, or gross domestic product.
An example, showing the Electoral College on a cartogram, is the diagram in
Fig.\,\ref{electoral_college}B which we adapted from
Wikipedia~\cite{WikipediaCart}.
Similar cartograms have been shown in the news
media~\cite{WashingtonPost16, NYTimes16}.
Here each elector is represented by a small square.
The squares are then positioned with two objectives in mind.
First, the shapes on the cartogram should resemble those on the map in
Fig.\,\ref{electoral_college}A. 
Second, the set of neighboring states in
Fig.\,\ref{electoral_college}A and Fig.\,\ref{electoral_college}B
should be the same.
Satisfying both objectives is not trivial.
A careful comparison with Fig.\,\ref{electoral_college}A shows that,
for example, Arizona (AZ) and Texas (TX) incorrectly appear as neighbors in
Fig.\,\ref{electoral_college}B.
On the other hand, the geographic neighbors Colorado (CO) and
Nebraska (NE) have been separated in Fig.\,\ref{electoral_college}B to
make space for other states in the vicinity.

For certain applications, it is perfectly acceptable that neighboring
states are split apart.
So long as the areas of the states are proportional to the number of
electors, such representations are called noncontiguous
cartograms~\cite{Olson76}.
Dorling's circular cartograms are good examples of noncontiguous
cartograms that, while not strictly maintaining the topology,
indicate where the represented regions are located~\cite{Dorling96}.
Contiguous cartograms, by contrast, not only rescale the regions, but also keep the
topology intact (i.e., neighbors on the map are neighbors on the
cartogram and vice versa).

The methods that have been proposed for generating contiguous
cartograms fall into two distinct categories.
The first group consists of algorithms that operate only on the boundaries of
regions~\cite{Dougenik_etal85, Merrill_etal92, HouseKocmoud98, Keim_etal04,
  Keim_etal05, InoueShimizu06, Kamper_etal13, Cano_etal15, Sagar14}. 
Each region is represented by one or multiple polygons.
The input to these algorithms are a finite number of polygon corners
$(x_1,y_1),\ldots,(x_n,y_n)$.
Here $(x_i,y_i)$ is a projection of the longitude and latitude,
usually obtained from a conventional projection (e.g., plate carr\'ee
or an equal-area projection).
The algorithm generates transformed polygon coordinates
$\mathbf{T}(x_1,y_1),\ldots,\mathbf{T}(x_n,y_n)$.
For the first group of algorithms, these $n$ points are in fact the
only output and, hence, we refer to them as ``boundaries-only''
  algorithms.
In other words, boundaries-only methods do not transform points
that are in the interior of a polygon.
For example, on a US state cartogram (such as
Fig.\,\ref{electoral_college}B) we would not be able to uniquely
locate a state capital such as Austin, TX, because it is far from any
state border.
One might symbolically place all capitals at the centroid of the corresponding polygon, but some centroids might be outside the polygon if it is concave or contains holes (e.g., lakes or enclaves).
The situation is even more complicated if we want to represent multiple distinct points or lines (e.g., rivers or roads) inside a state as distinct objects on a boundaries-only cartogram.

The second group of contiguous cartogram algorithms approaches the problem from a different point of view by producing a continuous transformation $\mathbf{T}$ for the entire
continuous set of longitudes and latitudes on the input map, including
coordinates that are not on a boundary~\cite{Tobler73, Cauvin_etal89,
  GuseinTikunov93, EdelsbrunnerWaupotitsch97, GastnerNewman04, Sun13}.
We refer to this group as ``all-coordinates'' algorithms.
Generating the map projection $\mathbf{T}$ for all longitudes and
latitudes can be computationally more demanding than only shifting the
boundary coordinates.
In fact, for applications where only the boundaries are of interest -- as is
the case for the US election map -- the boundaries-only
algorithms can give adequate results.
However, the run time of these discrete algorithms typically increases
steeply with the number of corners.
As a result, they often rely on coarse-grained input
to gain speed, for example by removing Michigan's Upper Peninsula from
the US map~\cite{Keim_etal04, Keim_etal05, Cano_etal15}.
If we wish to show data that are resolved at a scale much
finer than the polygons to be displayed [e.g., graticules for a fine,
spatially regular grid~\cite{Hennig13} or individual addresses], the
all-coordinates algorithms usually outpace their
boundaries-only counterparts.

\begin{figure*}
  \centering
  \includegraphics[width=\linewidth]{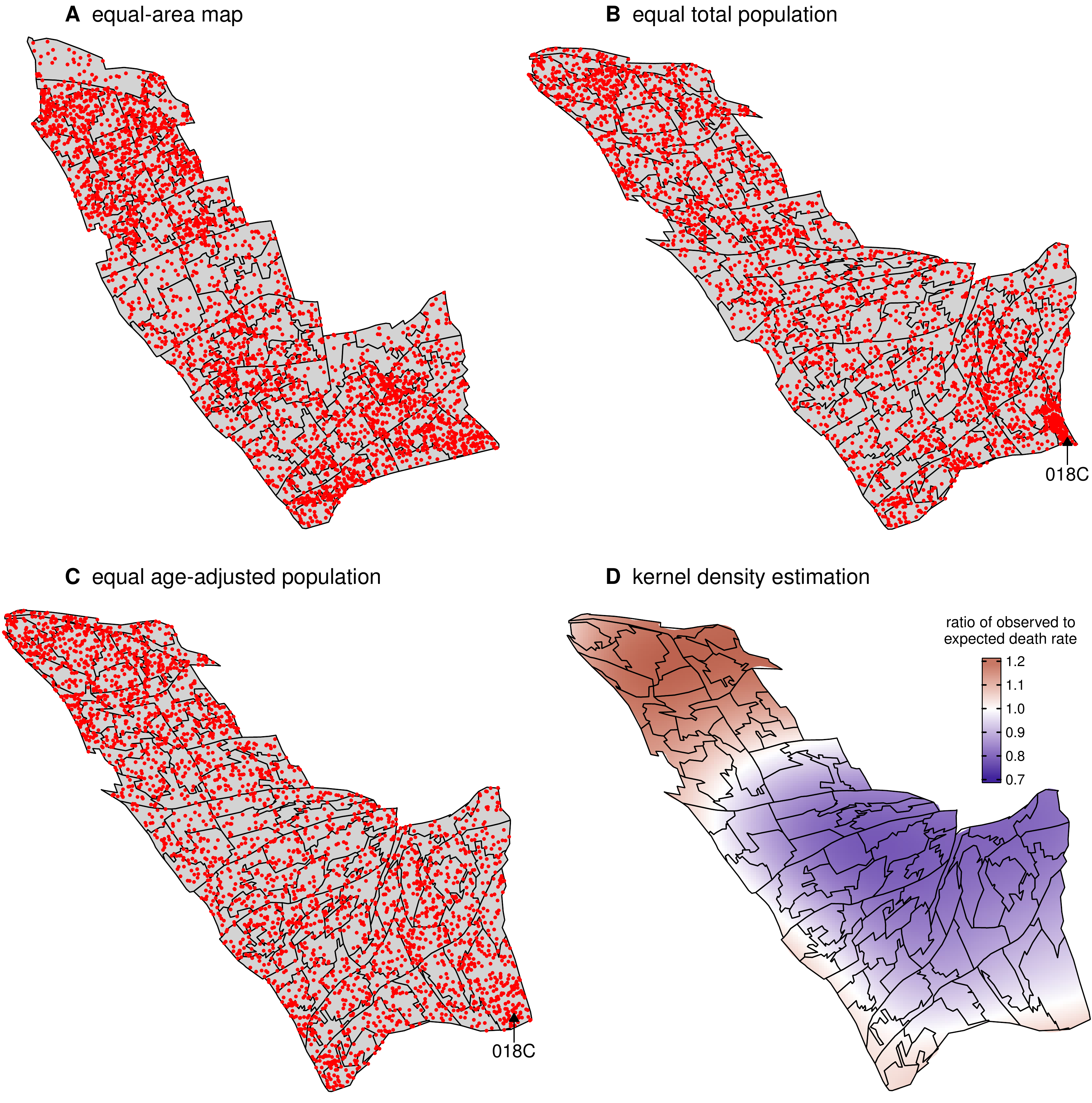}
  \caption{
    Maps with scatter plots of death cases in Kensington and Chelsea
    between 2011 and 2014 on (A) an equal-area map and on cartograms
    equalizing (B) the total population in each Lower Layer Super
    Output Area (LSOA) and (C) age-adjusted population (i.e., the
    expected number of deaths given the age and gender composition of the LSOA). 
    Cartogram (B) reveals a high per-capita mortality in LSOA 018C in the southeast of the borough
    caused by a nursing home located inside this polygon. 
    When accounting for the heterogeneous age distribution across the
    borough in (C), LSOA 018C has approximately the expected number of
    death cases.
    In other LSOAs, however, the expected and observed numbers differ.
    A kernel density estimate in panel (D) indicates an increasing trend in the
    age-adjusted death rate from the southeast to the northwest.
  }
  \label{london_fig}
\end{figure*}

In this article, we describe an all-coordinates algorithm that only
needs a few seconds to produce the complete projection $\mathbf{T}$
for realistic input.
Knowing $\mathbf{T}$ will allow us to show the positions of all US
state capitals with respect to the states' boundaries
(Fig.\,\ref{diff_and_new}B) and the coordinates of individual death
cases in London (Fig.\,\ref{london_fig}B and C).

\section*{Previous all-coordinates methods to produce a cartogram
  projection}

For the sake of concreteness, let us assume that we want to
make a cartogram whose areas are proportional to the population.
We define the population density as the function $\rho(x,y)$
such that a small rectangular area element with the corners $(x\pm
dx/2, y\pm dy/2)$  contains the population $\rho(x,y)\,dx\,dy$.
Some data allow us to model $\rho(x,y)$ with variations on fine
spatial scales. (Our application below to the mortality statistics of Kensington and Chelsea belongs to this category.)
In other cases, it is more natural to model $\rho(x,y)$ as a piecewise
constant function.
For example, California's 55 electors can be represented by a constant
density in this state equal to the number of electors divided by the
state's geographic area.

An accurate cartogram must project the rectangle $(x\pm
dx/2, y\pm dy/2)$ onto a quadrilateral $\mathbf{T}(x\pm dx/2, y\pm
dy/2)$ in such a way that the area of the quadrilateral is
proportional to $\rho\,dx\,dy$. 
In other words, we are looking for a two-dimensional function
$\mathbf{T}=(T_x, T_y)$ such that $\rho(x,y)\,dx\,dy =
\bar{\rho}\,dT_x\,dT_y$ where $\bar{\rho}$ depends neither on $x$ nor
$y$.
Such a transformation $\mathbf{T}$ is called a
  density-equalizing projection.
Taking the limits $dx\to 0$ and  $dy\to 0$ and assuming that
$\mathbf{T}$ is differentiable, we obtain the
condition~\cite{Tobler73, GastnerNewman04},
\begin{equation}
 \frac{\partial T_x}{\partial x}\frac{\partial T_y}{\partial y} -
 \frac{\partial T_x}{\partial y}\frac{\partial T_y}{\partial x} =
 \frac{\rho(x,y)}{\bar{\rho}}\ ,
  \label{detJ}
\end{equation}
which is called a prescribed Jacobian equation
\cite{dacorogna1990partial, avinyo2003maps}.
For convenience, we choose the constant $\bar{\rho}$ to be the
spatially averaged density so that the total mapped area is preserved.

Equation \ref{detJ} alone does not uniquely specify $\mathbf{T}$
because it is only one single equation for the two unknowns $T_x$ and
$T_y$~\cite{GuseinTikunov93}.
As a consequence, there are infinitely many different strategies to obtain a
density-equalizing projection $\mathbf{T}$.
In practice, however, only a few methods are computationally efficient,
produce attractive graphics and are independent of the choice of
coordinate axes.
Most of the methods that have been proposed in the literature are based on
physical analogies.
A common metaphor is to view the undistorted input map as a rubber
sheet.
Forces or stresses act on the rubber sheet such that the points move
toward equilibrium positions that satisfy
Eq.\,\ref{detJ}~\cite{Cauvin_etal89, Sun13}.
Although such mechanical metaphors make intuitive sense, there is no
direct physical connection between force and area. 
Therefore, it is not immediately obvious how the forces should be
chosen as functions of $\rho(x,y)$ to ensure that Eq.\,\ref{detJ} is valid.
Some methods treat the term ``force'' in a less literal sense so that
the area constraints are more explicitly part of the
equations~\cite{Dougenik_etal85, HouseKocmoud98}.
However, these algorithms must take special care to avoid topological
errors (e.g., regions that are flipped or boundaries that intersect
themselves) during the relaxation of the forces.
Another method, based on neural networks, starts by placing sample
points on a regular grid~\cite{Henriques:2009:CCC:1552426.1552450}.
During the training of the network, the samples are attracted toward
regions of high density to mimic the population distribution. 
Their final positions define a mapping which can produce a cartogram
by considering its inverse. 
However, a large number of sample points is necessary to produce
smooth boundaries.

An alternative physical metaphor is to view the process that generates
the cartogram as the flow of a fluid.
In this analogy, we think of the map as a Petri dish covered with a thin
layer of water.
In an experiment, we would model the population density $\rho(x,y)$ by
injecting small particles with spatially varying concentrations into
the water layer.
The particles then diffuse across the entire Petri dish.
In the long run, the probability density function of finding a particle
becomes a constant everywhere inside the dish.
We can make a cartogram by translating this simple physical model of
density equalization into a geographic map projection.

\begin{figure}
\centering
\includegraphics[width=\linewidth]{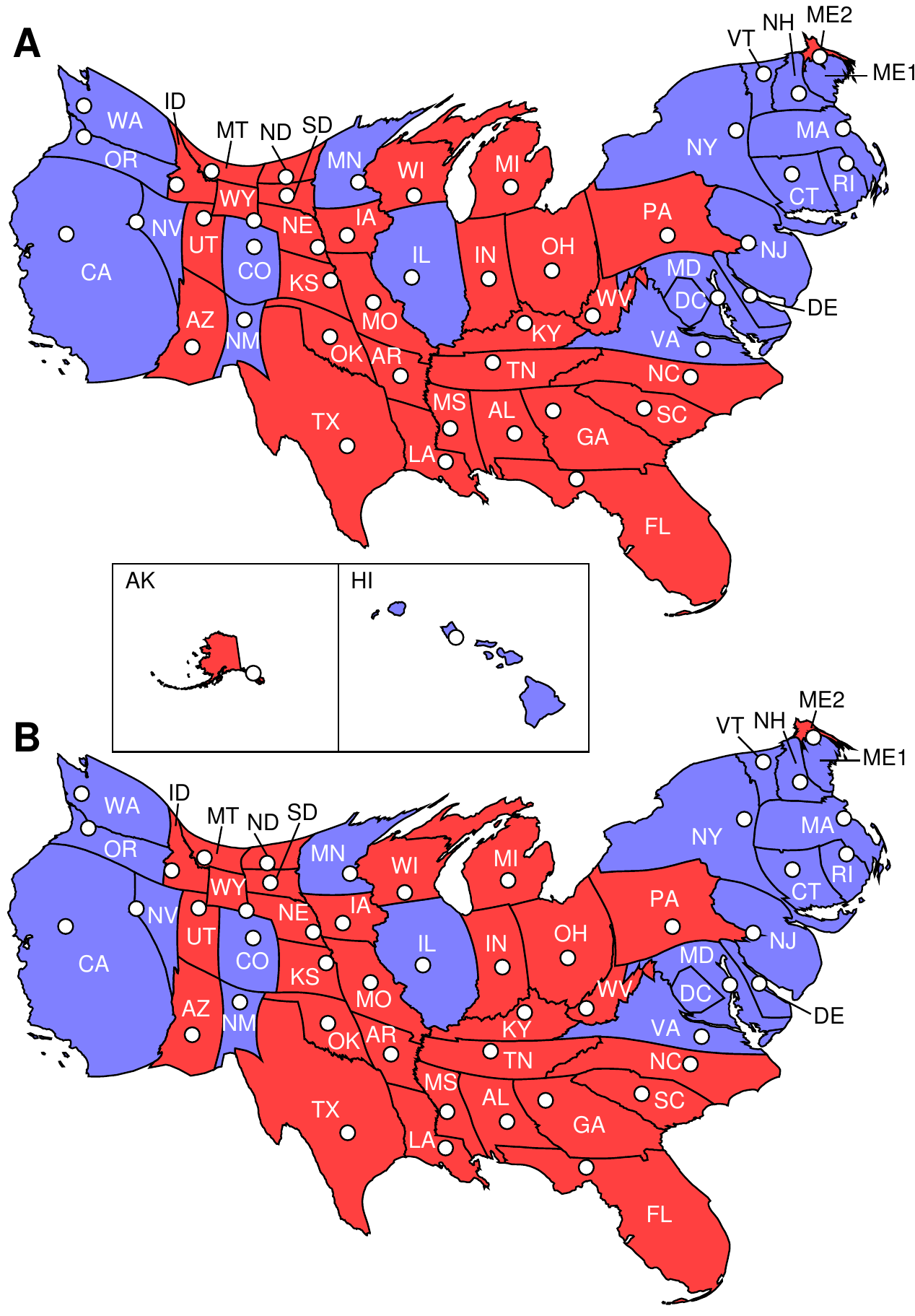}
\caption{
     The 2016 US Electoral College vote represented on cartograms generated with
    (A) the diffusion algorithm of Ref.\,\cite{GastnerNewman04} and (B)
    the alternative flow-based algorithm based on Eq.~\ref{rho_ffb}--\ref{rhotilde_mn}.
    The insets for Hawaii and Alaska apply to both (A) and (B) as
    these regions' areas match both cartograms.
    All areas differ by $<$1\% from their target values (i.e., the
    proportion of votes in the Electoral College). 
    Cartograms (A) and (B) differ in detail, but appear remarkably similar
    considering that generating (B) needs only 2.5\% of the time required by the
    diffusion algorithm.
    The white circles indicate the positions of the state capitals.
}
\label{diff_and_new}
\end{figure}

The most familiar process that equilibrates the density is Brownian
motion.
On a macroscopic scale, the Fokker-Planck equation that describes
Brownian motion is Fick's second law
$\partial\rho / \partial t = D\nabla^2\rho$.
Here $t$ stands for time, $D$ is the diffusivity and $\nabla^2$ is the
Laplace differential operator.
This equation, known as the diffusion or heat equation, is at the
heart of the ``diffusion cartogram'' method~\cite{GastnerNewman04, avinyo2003maps}.
An example of a diffusion cartogram is Fig.\,\ref{diff_and_new}A where
we show the US Electoral College results. 
The diffusion algorithm guarantees that, unlike in
Fig.\,\ref{electoral_college}B, each state keeps its neighbors while
still reaching the target areas to any desired level of accuracy.
The diffusion cartogram distorts the shapes of the states, which is
inevitable for any contiguous cartogram method. 
The shapes are, however, still recognizable; this is one of the
reasons why diffusion cartograms have become popular in the past
decade~\cite{NusratKobourov16}.
Another reason is that, despite the apparent complexity of the
equations, they can be computed relatively efficiently.

However, Fickian diffusion is only one of many types of fluid
dynamic rules that make particle densities equal everywhere.
As we argue now, there is an alternative that is computationally
more efficient while producing cartograms of comparable quality.

\section*{Flow-based cartogram with linear equalization}

In a flow-based cartogram, the population density $\rho$ is treated
not only as a function of position $\mathbf{r} = (x,y)$, but also as a function of 
time $t$.
For a density-equalizing projection, the density must approach its mean in
the long run: $\lim_{t\to\infty}\rho(x,y,t) = \bar{\rho}$
for all $x$ and $y$.
That is, the particles must flow in such a way that all initial differences
in their density are completely leveled out over time.
This condition alone, however, does not yet define the projection
$\mathbf{T}$.
 We must also know the velocity $\mathbf{v}(x,y,t)$ with which a point at
 $(x,y)$ is dragged along by the flow at time $t$. 
Because there are no sinks or sources in the flow, $\mathbf{v}$ must
satisfy the mass conservation equation, also known as the continuity equation,
\begin{equation}
	\label{eq:continuity_equation}
	\frac{\partial\rho}{\partial t} + \nabla \cdot (\rho \mathbf{v}) = 0.
\end{equation}
If we know $\mathbf{v}(x,y,t)$ for all $x$, $y$, and $t$, we can
compute the position $\mathbf{r}(t)$ for a point that is initially at
$\mathbf{r}(0)$,
\begin{equation}
  \mathbf{r}(t) = \mathbf{r}(0) + \int_0^t\mathbf{v}(\mathbf{r}(t'), t')dt'.
  \label{rt_diff}
\end{equation}
The projection $\mathbf{T}$ is the function that shifts
$\mathbf{r}(0)$ to $\lim_{t\to\infty}\mathbf{r}(t)$. 
In the SI Appendix (section 2), we explain in more detail why $\mathbf{T}$ is
density-equalizing.

We can satisfy the continuity equation while simultaneously demanding Fick's law $\mathbf{v} =
-D(\nabla\rho)/\rho$.
Substituting Fick's law into Eq.~\ref{eq:continuity_equation} shows
that the evolution of $\rho$ is then governed by the heat equation
$\partial\rho / \partial t = D\nabla^2\rho$.
This is the key motivation behind the diffusion cartogram method~\cite{GastnerNewman04}, but Fickian diffusion is only one special case among a large class of
processes in which $\rho$ relaxes to its mean density while satisfying 
the continuity equation for some velocity field $\mathbf{v}$. 
One advantage of Fickian diffusion is that the corresponding flow is
guaranteed to be free of vortices that could cause severe local
distortions in $\mathbf{T}$.
However, Fickian diffusion is not unique in this respect (see section
2 of the SI
Appendix) so that one is left wondering whether other vortex-free,
mass-conserving processes might also be suitable for generating
cartograms.
As we now argue, if we replace the heat equation by a linear equalization of the density toward the mean,
\begin{equation}
  \rho(x,y,t) = 
  \begin{cases}
    (1-t)\,\rho_0(x,y) + t\bar{\rho} & \text{if $t\leq1$},\\
    \bar\rho & \text{if $t>1$},
  \end{cases}
  \label{rho_ffb}
\end{equation}
we can indeed compute $\mathbf{T}$ significantly faster.
It has been shown that there exists a velocity field $\mathbf{v}$ for
Eq.~\ref{rho_ffb}  so that the resulting transformation $\mathbf{T}$
satisfies Eq.~\ref{detJ}~\cite{dacorogna1990partial}.
We derive the concrete formulas for $\mathbf{v}$ in the SI Appendix
(section 2) and
only give a brief summary here.

After an affine transformation of all coordinates, we place the mapped
area inside a rectangular box with bounding coordinates
$x_\text{min}=0$, $x_\text{max}=L_x$, $y_\text{min}=0$,
$y_\text{max}=L_y$.
(For later convenience, we choose $L_x$ and $L_y$ to be integers.)
If we demand that there is no flow through the edges of the box, the
velocity for $t\leq 1$ can be expressed in terms of sine and cosine
Fourier transforms,
\begin{align}
  v_x(x,y,t) = -\frac{L_y}{\pi\rho(x,y,t)} &\sum_{m=1}^\infty \sum_{n=0}^\infty
  \bigg[\frac m{m^2L_y^2+n^2L_x^2} \tilde\rho_{mn}
  \nonumber\\
  &\times\sin\left(\frac{m\pi
  x}{L_x}\right) \cos\left(\frac{n\pi y}{L_y}\right)\bigg]\ ,
  \label{vx_cable}\\
  v_y(x,y,t) = -\frac{L_x}{\pi\rho(x,y,t)} &\sum_{m=0}^\infty \sum_{n=1}^\infty \bigg[\frac
  n{m^2L_y^2+n^2L_x^2} \tilde\rho_{mn}
  \nonumber\\
  &\times\cos\left(\frac{m\pi
  x}{L_x}\right) \sin\left(\frac{n\pi y}{L_y}\right)\bigg]
  \label{vy_cable}
\end{align}
with
\begin{align}
  \tilde\rho_{mn} &= \frac 4 {(\delta_{m0}+1)(\delta_{n0}+1)}
  \label{rhotilde_mn}\\
  &\times\int_0^{L_x} \int_0^{L_y} \rho(x',y',0)\cos\left(\frac{m\pi
      x'} {L_x}\right) \cos\left(\frac{n\pi y'} {L_y}\right) dx'dy'.
  \nonumber
\end{align}
Here $\delta_{00}=1$ and $\delta_{m0}=0$ if $m\neq0$.
For $t>1$, we simply obtain $v_x(x,y,t)=v_y(x,y,t)=0$.

Equations \ref{vx_cable}--\ref{rhotilde_mn} look superficially
similar to the corresponding equations in the diffusion-based
cartogram~\cite{GastnerNewman04}, but there are two important
differences.
First, neither the sums in Eq.\,\ref{vx_cable}, \ref{vy_cable} nor the integral
in Eq.\,\ref{rhotilde_mn} depend on $t$ so that the Fourier
transforms need to be computed only once at the beginning of the
calculation.
Second, after we have computed the Fourier transforms, here we only
require quick arithmetic operations: addition, subtraction,
multiplication, and division.
For a diffusion cartogram, by contrast, we must repeatedly calculate
time-dependent Fourier transforms and evaluate the exponential
function during the integration of Eq.\,\ref{rt_diff} (see section 2
of the SI Appendix).
The speed of computing the exponential function depends on details of
the implementation and hardware, 
but is in general much slower than addition, subtraction,
multiplication, or division~\cite{Brent75}.

These mathematical differences alone already cut the time needed per
integration step by more than half.
Another simplification compared with the diffusion cartogram is that we
need to integrate Eq.\,\ref{rt_diff} only until $t=1$ instead of
$t=\infty$.
The benefit is that we no longer need to check whether the improper integral over
the velocity has sufficiently converged.
Most importantly, however, the integrals from different starting
points $\mathbf{r}(0)$ can be performed in parallel as we now explain.

We overlay the map with an $L_x\times L_y$ square grid.
For these $L_xL_y$ coordinates, we compute the sums and integrals in
Eq.\,\ref{vx_cable}--\ref{rhotilde_mn} at the start of the calculation
with the fast Fourier transform algorithm~\cite{FrigoJohnson05}.
We have found that the time needed for this one-time procedure is a
negligible fraction of the total run time. 
After storing the $L_xL_y$ Fourier transforms in memory, we obtain
$\mathbf{v}(\mathbf{r},t)$ at
each grid point $\mathbf{r}$ with basic arithmetic.
Subsequently, we find the integrand in Eq.\,\ref{rt_diff} for non-grid
positions $\mathbf{r}$ by interpolating between the grid points.
We numerically approximate the integral using a predictor-corrector
method that automatically adapts the size of the next time step.
During each step, we distribute the integration of the $L_x L_y$
distinct integrands to different processing units.
In practice, given the wide availability of multi-core processors nowadays, this
parallelization enormously boosts the speed of the calculation.

\section*{Benchmarking the algorithm with data for the USA, India, and China}

\begin{figure*}
  \centering
  \includegraphics[width=\linewidth]{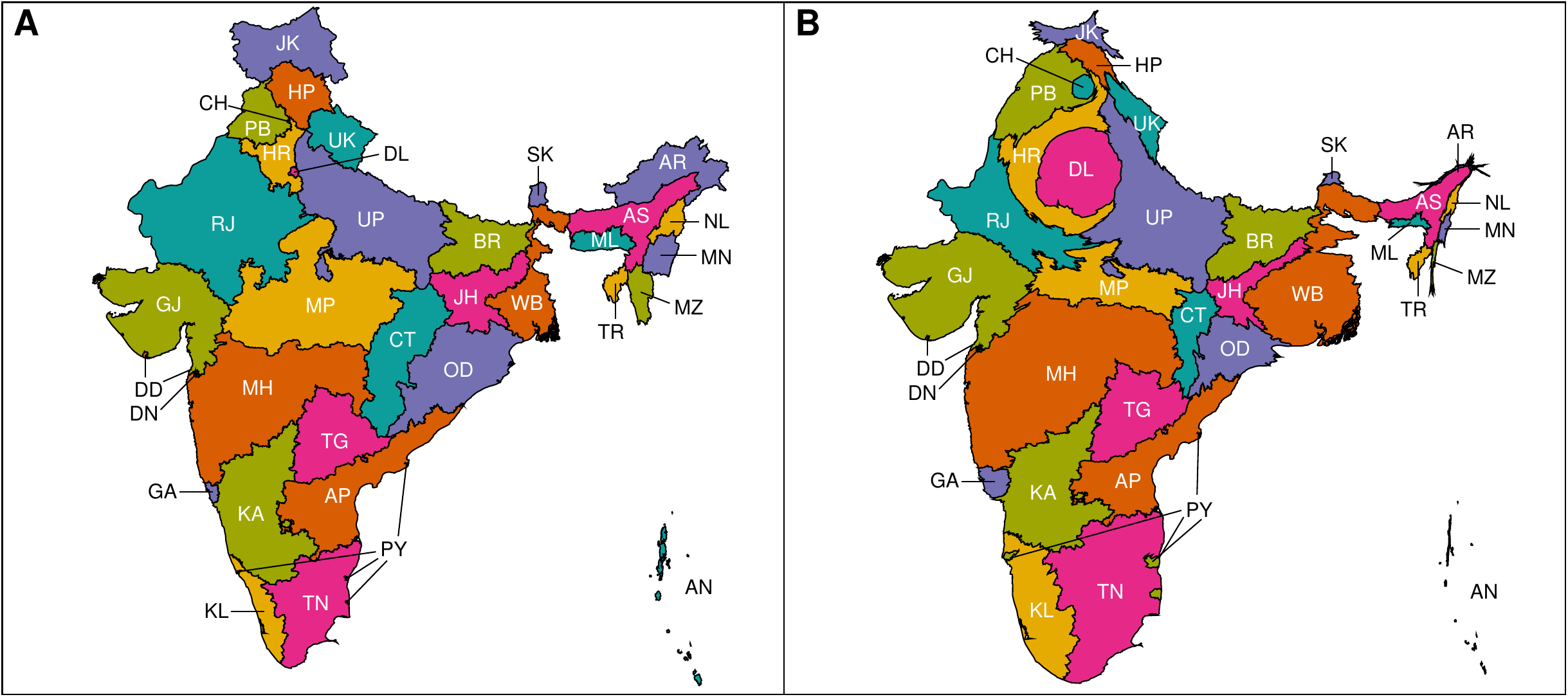}
  \caption{
    The states and union territories of India on (A) an equal-area map,
    (B) a cartogram where the area of each region is proportional to
    GDP (data from Statistics Times~\cite{StatTimes17}).
    The two largest states by area, Rajasthan (RJ) and Madhya Pradesh
    (MP), shrink on the cartogram because they only rank 7th and 10th
    in GDP, respectively. 
    Maharashtra (MH), the state with the highest GDP, slightly grows
    on the cartogram. 
    Even more striking is the increase of Delhi (DL): although small in
    area, the capital city has a higher GDP than many larger states.
    The opposite happens for Arunachal Pradesh (AR) and several other
    northeastern states because they rank low in GDP.
    Our algorithm only needs 2.6 seconds to construct the cartogram.
AN, Andaman and Nicobar Islands;
AP, Andhra Pradesh;
AS, Assam;
BR, Bihar;
CH, Chandigarh;
CT, Chhattisgarh;
DN, Dadra and Nagar Haveli;
DD, Daman and Diu;
GA, Goa;
GJ, Gujarat;
HR, Haryana;
HP, Himachal Pradesh;
JK, Jammu and Kashmir;
JH, Jharkhand;
KA, Karnataka;
KL, Kerala;
MN, Manipur;
ML, Meghalaya;
MZ, Mizoram;
NL, Nagaland;
OD, Odisha;
PY, Puducherry;
PB, Punjab;
RJ, Rajasthan;
SK, Sikkim;
TN, Tamil Nadu;
TG, Telangana;
TR, Tripura;
UP, Uttar Pradesh;
UK, Uttarakhand;
WB, West Bengal.
  }
  \label{india}
\end{figure*}

\begin{figure*}
  \centering
  \includegraphics[width=\linewidth]{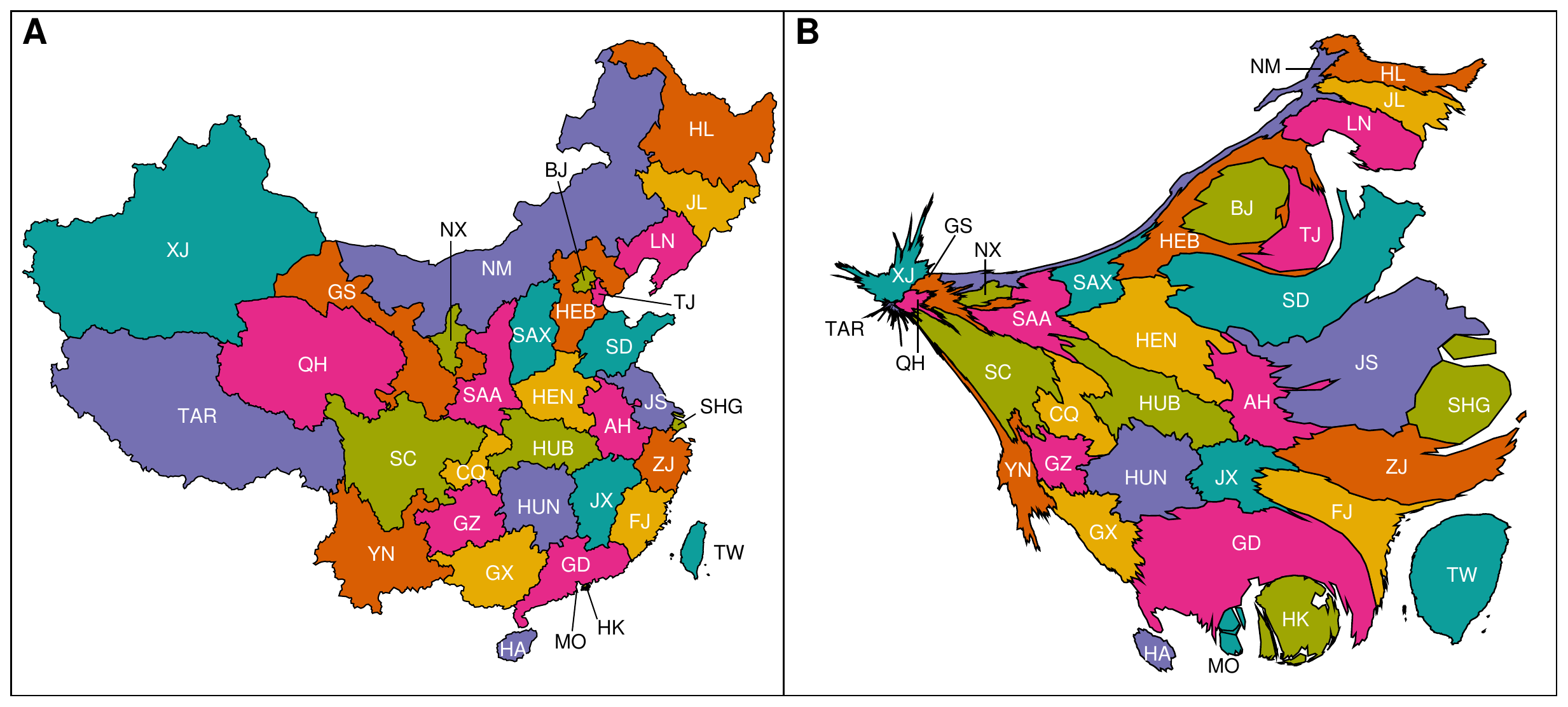}
  \caption{
    Provincial-level administrative divisions of mainland China and Taiwan on (A) an
    equal-area map, (B) a cartogram where areas are proportional to
    GDP (data from Wikipedia~\cite{WikipediaChina}).
    Some coastal cities such as Shanghai (SHG) and Hong Kong (HK)
    increase remarkably on the cartogram.
    By contrast, western states such as Xinjiang (XJ) and the Tibet Autonomous Region (TAR) shrink dramatically.
    Despite the substantial deformations, our algorithm only needs 2.7
    seconds to construct the cartogram.
AH, Anhui;
BJ, Beijing;
CQ, Chongqing;
FJ, Fujian;
GS, Gansu;
GD, Guangdong;
GX, Guangxi;
GZ, Guizhou;
HA, Hainan;
HEB, Hebei;
HL, Heilongjiang;
HEN, Henan;
HUB, Hubei;
HUN, Hunan;
NM, Inner Mongolia;
JS, Jiangsu;
JX, Jiangxi;
JL, Jilin;
LN, Liaoning;
MO, Macao;
NX, Ningxia;
QH, Qinghai;
SAA, Shaanxi;
SD, Shandong;
SAX Shanxi;
SC, Sichuan;
TW, Taiwan;
TJ, Tianjin;
YN, Yunnan;
ZJ, Zhejiang.
  }
  \label{china}
\end{figure*}

We have implemented the algorithm based on Eq.\,\ref{rho_ffb}--\ref{rhotilde_mn} as a C
program. 
In this section, we illustrate its performance with three case studies: the 2016 vote
in the US Electoral College (Fig.\,\ref{diff_and_new}B), the
distribution of India's gross domestic product (GDP) by state (Fig.\,\ref{india}),
and mainland China's and Taiwan's GDP by province (Fig.\,\ref{china}).

In each case, we first project the longitudes and latitudes of the
territorial borders with an Albers equal-area conic projection onto
a flat two-dimensional space. 
As described above, we embed the resulting map (Fig.\,\ref{electoral_college}A,
\ref{india}A and \ref{china}A, respectively) inside an
$L_x\times L_y$ rectangle whose edges act as reflecting boundaries for the
flow.
The rectangular box should, on one hand, be chosen large
enough so that the cartogram is independent of the boundary
conditions.
On the other hand, it should not be so large that we spend the bulk of the
run time on computing the projection $\mathbf{T}$ far from
the region of interest.
As a compromise, we have chosen the side length equal to 1.5 times the
maximum of the countries' north-south and east-west extent. (These
rectangular boxes are larger than the frames shown in Fig.\,\ref{india} and
Fig.\,\ref{china} whose purpose is purely to visually separate the
different panels in the figure.)
The space between the country and the edges of the box is filled with
the mean density $\bar\rho$.
Other choices are conceivable and may improve shape preservation (e.g., by more faithfully retaining the outer boundaries of the map), but they
would result in more complex computer code.

For the discrete Fourier transforms, we divide the large rectangular box
into a grid of $L_x\times L_y$ smaller squares (in our examples $L_x = L_y = 512$, but the number can be adjusted if necessary)  whose sizes are just fine enough
to discern the smallest geographic regions on each map: Washington, DC
in the USA; Daman and Diu in India (abbreviated by DD in
Fig.\,\ref{india}); and Macao in China (MO in Fig.\,\ref{china}).
Officially, these regions have neither the status of a state nor a
province: Washington DC is a district, Daman and Diu a union
territory, and Macao a Special Administrative Region.
We still include these regions on the cartograms because they are
typically included on maps showing the states and provinces of their
respective countries.\footnote{We exclude the island territory of
  Lakshadweep from the maps of India because it is so small that it is
  neither visible on an equal-area map nor on a GDP cartogram.}

When numerically integrating Eq.\,\ref{rt_diff}, the choice of time
steps determines how accurately we estimate $\mathbf{r}(1)$.
One possible strategy for achieving a highly accurate cartogram is to
take a large number of small steps.
After some experiments, we have decided to use a different strategy
that achieves quicker run times and ultimately also comes arbitrarily
close to a perfectly density-equalizing map.
We use only a moderate number of adaptive time steps ($\approx 100$ in a typical run; the exact number is determined at runtime)
during the initial integration.
We expedite the convergence by applying a Gaussian blur of moderate
width to the initial density prior to starting the integration.
After one round of integration, the areas do not yet perfectly
match their targets.
For example, Washington DC still needs to grow by a
factor $\approx 50$.
The key feature is to use the output of the first integration as input to
another round of integration, which then usually comes closer to the
objective areas.
By repeating the integration sufficiently often, we have in all test
cases observed that we can reach the objective areas with arbitrary
precision.
For the contiguous 48 states of the USA, we perform five iterations.
Afterward, even for the extreme case of Washington DC, the smallest
region in land area, the cartogram area differs by only $0.31\%$ from the
objective area. 
For India we iterate the integration twelve times and for China six times.
The maximum differences between target and objective area are then
$0.72\%$ for the Andaman and Nicobar Islands (AN in
Fig.\,\ref{india}B) and $0.83\%$ for Tibet (TAR in
Fig\,\ref{china}B), respectively.
These differences are certainly so small that they cannot be
detected by eye.
We generally set a maximum relative area error of $<1\%$, defined as 
\[
\text{relative area error} = \frac{\text{target area}-\text{objective area}}{\text{objective area}}\ ,
\]
as stopping criterion for the algorithm.

This level of accuracy is all the more remarkable when considering the
speed of our implementation.
On a Dell Precision\textsuperscript{\textregistered} T7810 workstation
with a 12-core Intel\textsuperscript{\textregistered}
Xeon\textsuperscript{\textregistered} E5-2680V3
processor and an Ubuntu 16.04.2 operating system, we need $1.5$
seconds for the US Electoral College cartogram
(Fig.\,\ref{diff_and_new}B), $2.6$ seconds for the India GDP 
cartogram (Fig.\,\ref{india}B), and $2.7$ seconds for the China
GDP cartogram (Fig.\,\ref{china}B).
Compared with the diffusion algorithm, which needs $59.5$ seconds to
generate the US cartogram (Fig.\,\ref{diff_and_new}A) with equal
accuracy, this is a speedup by roughly a factor $40$.
Among other all-coordinates cartogram algorithms, only the
rubbersheet method Carto3F~\cite{Sun13} can achieve comparable speed,
but not for all types of input.
For a cartogram of Chinese provinces,  Carto3F needs 8 minutes of computer time.
Our fast flow-based method achieves smaller area errors in a fraction of this time.

\section*{Benchmarking with data for mortality in Kensington and Chelsea (London) 2011--2014}
As noted above, cartogram algorithms that generate the complete
density-equalizing projection $\mathbf{T}$ are particularly
advantageous when displaying demographic data that are individual
points on a map.
We now demonstrate how our algorithm can be applied to such input and how we can use it to compare different statistical models.
The data also serve as another benchmark for the speed of our method.
Our example involves the locations of all $3197$ death cases in the London
borough of Kensington and Chelsea between the years 2011 and 2014.
The database from the UK's Office for National Statistics (ONS)~\cite{ONS16}
lists the number of deaths in each of London's 4835 Lower Layer Super
Output Areas (LSOAs).
A total of 103 LSOAs are located in Kensington and Chelsea.
We show the density of death cases in this borough on an equal-area
map in Fig.\,\ref{london_fig}A.
Each death corresponds to one point on the map placed at a random
position inside the LSOA where it occurred.

The point pattern on the equal-area map is spatially heterogeneous
with two bands of high density, one in the south and another in the
north, separated by a band of lower density in the middle. 
However, it remains unclear from the equal-area map whether the
differences in the spatial distribution of death cases are caused by
differences in per-capita mortality or by a heterogeneous population
density.
We can distinguish between these two effects by projecting the death
cases to a cartogram where each LSOA area is proportional to the
number of inhabitants (Fig.~\ref{london_fig}B).

The most striking feature on this cartogram is the high per-capita
mortality in the southeast corner of the borough.
The reason for the high number of death cases in the LSOA with the
ONS code ``Kensington and Chelsea 018C'' is a large proportion of
elderly, most likely because of the St.\ Wilfrid's nursing home
located in this LSOA.
Because mortality increases markedly as a person becomes elderly,
total population is too crude a measure to predict death rates.
We now show how to improve the prediction by using each LSOA's age-adjusted
mortality as the basis of a cartogram instead of the simple per-capita mortality displayed in
Fig.~\ref{london_fig}B.

Data from the ONS~\cite{ONS15, ONS16} include population size and
death cases in the following age groups for each LSOA: 0 years old,
1-4, 5-9, ..., 85-89, and $\geq 90$ years old, with each age group
divided into men and women.
For each of these 40 demographic subgroups, we can compute its total
mortality in western central London (i.e., Kensington and Chelsea as
well as the adjacent boroughs Brent, Westminster, Wandsworth, Hammersmith and
Fulham).
We denote by $p_j$ the size of the population that lives in this part
of London and belongs, because of its gender and age, to the
demographic group $j$. 
If there were $d_j$ deaths in this subpopulation, its region-wide per-capita
mortality is $m_j = d_j/p_j$.
The expected number of deaths in the $i$-th LSOA is thus $e_i = \sum_ j
p_{ij}m_j$, where $p_{ij}$ is the population that lives in LSOA $i$
and belongs to the demographic group $j$.
This approach is known in the public health literature as
age-adjustment~\cite{LilienfeldStolley94}.
Unlike the unadjusted population size $\sum_j p_{ij}$, the expected
value $e_i$ makes a fair comparison between, for example, an LSOA
mostly inhabited by a younger population and an LSOA with a large
proportion of elderly inhabitants such as 018C.

In Fig.\,\ref{london_fig}C, we show a cartogram with LSOA areas
proportional to $e_i$.
On this cartogram, the density of points in 018C is near the average in the
borough, visualizing that age is indeed an important predictor for local death
rates.
Across the borough, however, differences between death rates still
remain despite age-adjustment.
We can quantify the deviation from spatial homogeneity, for example,
with the Hopkins statistic $H$~\cite{HopkinsSkellam54}, which is a number between $0$ and $1$. 
If a point pattern is caused by a homogeneous Poisson process (i.e.,
deaths are independent and equally likely everywhere), then the
expected value of $H$ equals $0.5$.
The more clustered the points are, the larger $H$ is.
We find $H=0.524$ (95\% confidence interval $[0.518, 0.530]$) in
Fig.\,\ref{london_fig}C, indicating that the data are inconsistent
with a homogeneous Poisson process.

We show a kernel density estimate of the underlying probability
distribution in Fig.\,\ref{london_fig}D.
We use a bivariate normal kernel with a bandwidth chosen according to
Ref.~\cite{VenablesRipley02}.
The figure reveals a minimum in the age-adjusted death rate in the
east of the borough and a maximum in the north.
Previous studies have argued that indicators of health (e.g., life expectancy) in different parts of
London are positively correlated to average household income~\cite{Dorling13}.
A choropleth map of deprivation in Kensington and Chelsea~\cite{Economist17} does indeed follow a strikingly similar regional
pattern as the death rate in Fig.\,\ref{london_fig}D.

The flow-based method of Eq.\,\ref{rho_ffb}--\ref{rhotilde_mn} calculates the cartograms in Fig.\,\ref{london_fig}B and C in $1.6$ and $1.9$ seconds respectively.
To avoid boundary effects in Fig.\,\ref{london_fig}D, we also include
data for Kensington and Chelsea's neighboring boroughs when computing
the cartograms and the kernel density estimate.
The equivalent calculations with the diffusion-based method take $69.9$
and $99.5$ seconds respectively.

\section*{Measures of distortion}

Our algorithm is not only accurate and fast, but also generates
cartograms whose visual appearance is on par with previous methods. 
In Fig.\,\ref{diff_and_new} we directly compare the diffusion
cartogram of the USA (panel A) with the faster method based on Eq.\,\ref{rho_ffb}--\ref{rhotilde_mn} (panel B).
The border between Illinois (IL) and Indiana (IN) is straighter in Fig.\,\ref{diff_and_new}A than in Fig.\,\ref{diff_and_new}B and thus more similar to the input map (Fig.\,\ref{electoral_college}A).
On the other hand, the border between New Mexico (NM) and Colorado (CO) is straighter and Oklahoma's (OK) panhandle less bent in Fig.\,\ref{diff_and_new}B.
Overall, however, the differences between both cartograms are only subtle.

Because visual appearance is not a fully satisfactory criterion, we now turn to quantitative measures of distortion.
One way to compare the local distortion of different projections is by analyzing the Tissot indicatrix that is constructed as follows. 
Suppose we draw an infinitesimal circle at the coordinates $(x,y)$ on the input map.
Locally, the effect of the projection $\mathbf{T}$ is to deform the circle into an ellipse, called the Tissot indicatrix of $\mathbf{T}$ at $(x,y)$.
Figure~\ref{tissot} in the SI Appendix shows concrete examples of Tissot indicatrices for our benchmarking examples.
We denote the semi-major and -minor axes of the Tissot indicatrix by $a(x,y)$ and $b(x,y)$ respectively.
Two measures of the local distortion error are~\cite{Papadopoulos17}
\[
e(x,y) = \ln\left(\frac{a(x,y)}{b(x,y)}\right)
\]
and~\cite{snyder1987map}
\[
\tilde e(x,y) = 2\arcsin\left(\frac{a(x,y)-b(x,y)}{a(x,y)+b(x,y)}\right)\ .
\]
For a conformal (i.e., angle-preserving) map, we would have $a=b$ for
all $(x,y)$ and thus $e = \tilde e = 0$. This scenario would be ideal,
but, as we review in the SI Appendix (section 3), except in a few special cases there cannot be a conformal density-equalizing projection~\cite{GuseinTikunov93}.
As a global measure for the deviation of a cartogram from conformality, we can use for example either the spatially averaged or the maximum local distortion error,
\[
e_a = \frac{1}{|\Omega|}\int_\Omega e(x,y)dx\,dy, \qquad
e_\infty = \underset{(x,y) \in \Omega}{\sup} e(x,y),
\]
where $\Omega$ is the spatial domain of the input map.
In our comparison of the diffusion and fast flow-based algorithm in Table~\ref{tab:distortion}, we choose $\Omega$ to be the rectangular $L_x\times L_y$ bounding box that contains the area to be mapped as described above.
By replacing $e$ with $\tilde e$, we obtain similar measures $\tilde e_a$ and $\tilde e_\infty$.

\begin{table*}[ht]
\caption{Measures of distortion applied to the diffusion algorithm and the flow-based algorithm using Eq.~\ref{rho_ffb}--\ref{rhotilde_mn}. Smaller values are highlighted in bold.}
\begin{center}
\begin{tabular}{|c|c|c|c|c|c|c|c|c|c|}
    \hline
    Map & Algorithm & $e_a$ & $e_\infty$ & $\tilde e_a$ & $\tilde e_\infty$ & $\alpha$ & $\delta$ & $\theta$ & run time (seconds)\\
    \hline
    \multirow{2}{*}{USA} & diffusion & {\bf 0.278} & {\bf 6.85} & {\bf 0.273} & {\bf 3.01} & {\bf 2.01} & {\bf 17.1} & {\bf 0.0388} & $59.5$\\
    & fast flow-based & $0.285$ & $7.06$ & $0.280$ & $3.02$ & $2.04$ & $17.4$ & $0.0435$ & {\bf 1.5}\\
    \hline
    \multirow{2}{*}{India} & diffusion & {\bf 0.190} & $3.95$ & {\bf 0.185} & $2.59$ & $2.45$ & {\bf 39.0} & {\bf 0.0281} & $113.0$\\ 
 & fast flow-based & $0.191$ & {\bf 3.18} & $0.187$ & {\bf 2.34} & {\bf 2.40} & $39.7$ & $0.0290$ & {\bf 2.6}\\
 \hline
  mainland China & diffusion & $0.590$ & {\bf 5.07} & $0.553$ & {\bf 2.83} & $2.33$ & {\bf 18.6} & {\bf 0.0849} & $178.5$\\ 
 and Taiwan & fast flow-based & {\bf 0.570} & $8.16$ & {\bf 0.530} & $3.07$ & {\bf 2.19} & $20.6$ & $0.103$ & {\bf 2.7}\\
  \hline
  Kensington \& & diffusion & {\bf 0.161} & {\bf 6.86} & {\bf 0.154} & {\bf 3.01} & {\bf 2.03} & {\bf 22.1} & {\bf 0.0589} & $99.5$\\ 
 Chelsea (age adj.) & fast flow-based & $0.163$ & $7.08$ & $0.156$ & $3.03$ & $2.20$ & $24.1$ & $0.0615$ & {\bf 1.9}\\
 \hline
\end{tabular}
\end{center}
\label{tab:distortion}
\end{table*}

When computing $e$ and $\tilde e$, we need to know $\mathbf{T}$ at each coordinate $(x,y)$ so that these measures can only be applied to all-coordinates cartograms.
Measures that aim to quantify the distortions also for other types of cartograms must instead rely on the polygons defining each region.
In Table~\ref{tab:distortion}, we include three such measures from Ref.~\cite{Alam_etal15}: the average aspect ratio $\alpha$, the Hamming distance $\delta$ and the relative position error $\theta$.
We provide details of their definition in the SI Appendix (section 4).
Briefly, the aspect ratio of a region is the ratio of the larger to the smaller side length of the bounding rectangle with minimum area, minimized over all possible rotations with respect to the coordinate axes.
The Hamming distance between two polygons is the area lying within exactly one of them~\cite{Skiena08}. For the measurement in Table~\ref{tab:distortion}, we rescale each polygon on the input map and the corresponding polygon on the cartogram so that they have equal area.
We then calculate the minimum Hamming distance between these two polygons by shifting one polygon with respect to the other.
We define $\delta$ as the sum of the minimum Hamming distances, where the summation is over all corresponding pairs of polygons.
For the relative position error, we compute the angle between the line connecting the centroids of two polygons on the input map and the line that connects the centroids of the corresponding two polygons on the cartogram.
We obtain $\theta$ by averaging over all pairs of polygons~\cite{heilmann2004recmap}.

Most measures listed in Table~\ref{tab:distortion} exhibit only small relative differences in the range of a few percent between the diffusion and fast flow-based method. 
Diffusion performs a little better in the majority of examples and measures, but there are also cases where the fast flow-based method produces a smaller error.
Considering the vastly different run times, the fast flow-based method is the better solution as a general-purpose algorithm for interactive applets.

\section*{Conclusion}

The scientific value of cartograms can go far beyond providing
mere entertainment, shock, or amusement.
As ``isodemographic maps'' they have been used for mapping diseases
and mortality for several decades~\cite{Selvin_etal84, Dorling98, Wieland29052007} in order to improve health services~\cite{lovett2014using}.
Arguably, the technical challenge of computing the map projection has
so far prevented more widespread use.
We accompany this article with C code available at
\texttt{https://github.com/Flow-Based-Cartograms/go\_cart} to
alleviate some of the challenge.
The code optionally produces the graticule of the inverse transformation so that features found on the cartogram can be identified in the original domain.
We reconstruct the original positions by first approximating $\mathbf{T}$ as a piecewise linear function and then computing its inverse.
The speed of the cartogram algorithm depends on the number of
processing units available to the user.
 If the calculation runs on a multi-core web server, users will be able
to take full advantage of the parallelized code at no cost.
We hope that in this form the algorithm will be accessible to a wider
audience.

\acknow{This research was supported by the European Commission
  (project number FP7-PEOPLE-2012-IEF 6-4564/2013).}

\showacknow 



\newpage

\normalsize

\section*{SI Appendix}

\section{Cartogram of the popular vote in the 2016 US presidential election}

US presidential elections are indirect: voters do not directly elect the president, but instead choose electors for their state who represent a presidential candidate in the Electoral College. 
The candidate with most votes in the Electoral College becomes the next president. 48 out of 50 states and Washington DC apply a winner-takes-all rule: the presidential candidate with the largest number of votes cast by the population in the state wins all of the state's electoral votes. 
The only exceptions are Maine and Nebraska. 
These two states apply the congressional district method: besides two electors for the state's aggregate winner, each congressional district chooses one elector for the candidate with most votes in this district.

The composition of the Electoral College does not need to be an accurate representation of the nationwide popular vote.
The predominant winner-takes-all rule gives an advantage to a candidate who wins many states with narrow margins even if the opponent may have won more votes in the population as a whole.
Furthermore, the number of votes in the Electoral College is not strictly proportional to state populations.
There is a small bias in favor of less populated states by guaranteeing every state a minimum of three electors.

Historically, the winner of the nationwide popular vote has usually also won the Electoral College.
However, the 2016 election was one of the exceptions. Hillary Clinton gained 48.2\% of the popular vote, Donald Trump only 46.1\%.
Nevertheless, Trump won the Electoral College by 304 to 227 votes.

\begin{figure}[tbhp!]
\centering
\includegraphics[width=\linewidth]{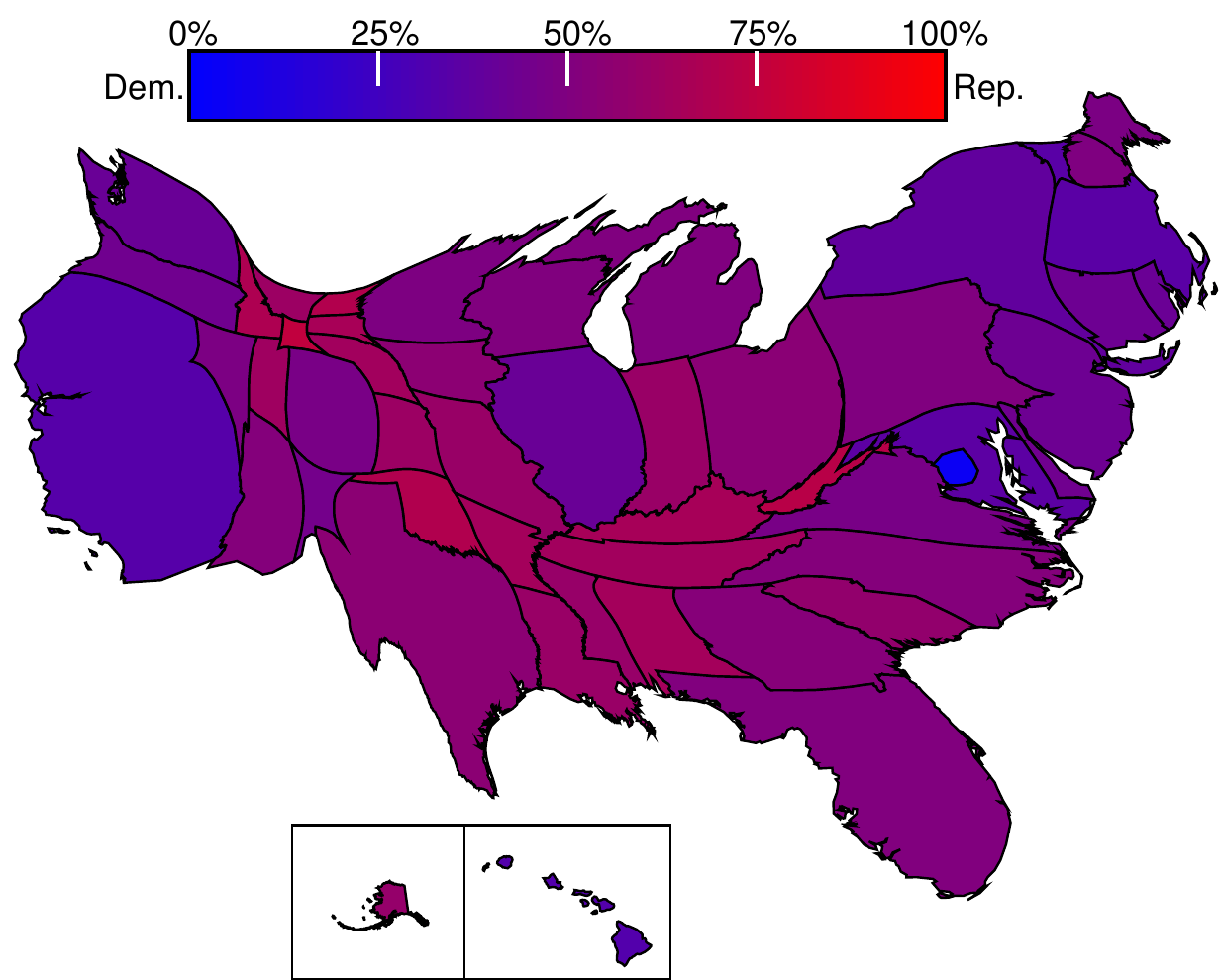}
\caption{The popular vote in the 2016 US presidential election on a cartogram made with the fast flow-based algorithm described in the main text.}
\label{uspopvote}
\end{figure}

We visualize the popular vote on a cartogram (Fig.~\ref{uspopvote}) by making each state's area proportional to the number of combined votes cast for Trump or Clinton in this state. We indicate the result with a color between blue (100\% for Clinton) and red (100\% for Trump). The shade of purple indicates how votes were split in each state. Cartograms with the same color scheme have been shown for previous US presidential elections~\cite{Gastner_etal05}.

\section{Motivating the equations used by the algorithm}

\subsection*{Flow-based density-equalizing projections}
Suppose we are given a population density $\rho_0(\mathbf{r})$ for
every point $\mathbf{r} = (x,y)$ in a rectangle defined by $0\leq
x\leq L_x$ and $0\leq y\leq L_y$.
Our objective is to map the rectangle onto itself with a
density-equalizing projection $\mathbf{T}$.
That is, assuming $\mathbf{T}$ is differentiable, it must satisfy
\begin{equation}
  \det(\nabla\mathbf{T}(\mathbf{r})) = \frac{\rho_0(\mathbf{r})}{\bar\rho}
  \label{det}
\end{equation}
for every point $\mathbf{r}$ in the rectangle.
The left-hand side is the Jacobian determinant
\[
\det(\nabla\mathbf{T}(\mathbf{r})) = \frac{\partial T_x}{\partial x}
\frac{\partial T_y}{\partial y} - \frac{\partial T_x}{\partial y}
\frac{\partial T_y}{\partial x}
\]
and the denominator in Eq.~\ref{det} is the spatially averaged density
\[
\bar\rho = \frac1{L_xL_y} \int_0^{L_x}\int_0^{L_y}
\rho_0(x,y)\,dx\,dy.
\] 
Loosely speaking, $\det(\nabla\mathbf{T}(\mathbf{r}))$ is the factor
by which a small area element near $\mathbf{r}$ is rescaled after
applying the transformation $\mathbf{T}$.
The general form of Eq.~\ref{det} is called a ``prescribed Jacobian
equation''.

The idea behind flow-based methods to find a solution $\mathbf{T}$ is
to define a sequence of densities $\rho(x,y,t)$, where the nonnegative
variable $t$ represents time.
We start from the given population density,
\begin{equation}
  \rho(x,y,0) = \rho_0(x,y),
  \label{initcond}
\end{equation}
and demand that $\rho$ approaches in the long run the spatially averaged
density,
\begin{equation}
  \lim_{t\to\infty}\rho(x,y,t) = \bar\rho\ .
  \label{lim_rho}
\end{equation}
For constructing a flow-based cartogram, we also need a
two-dimensional velocity field $\mathbf{v} = (v_x, v_y)$ for all $x$,
$y$, and $t$.
We define the map projection $\mathbf{T}_t$ of a point that
is initially at $\mathbf{r}=(x,y)$ by
\begin{equation}
  \mathbf T_t(\mathbf r) = \mathbf r + \int_0^t
  \mathbf v(\mathbf T_{t'}(\mathbf r)) dt'\ .
  \label{T_integral}
\end{equation}
We now argue that in the limit of infinite time, $\mathbf T_\infty$
is a density-equalizing projection (i.e., it satisfies Eq.~\ref{det})
if the combination of $\rho$ and $\mathbf v$ satisfies the continuity
equation
\begin{equation}
  \frac{\partial\rho}{\partial t} = - \nabla\cdot\mathbf J.
  \label{continuity}
\end{equation}
Here
\[
\mathbf J = \rho\mathbf v
\]
is the flux (i.e., the population that flows per unit time through a
line of unit length perpendicular to $\mathbf J$) and $\nabla\cdot\mathbf{J} =
\frac{\partial J_x}{\partial x}+\frac{\partial J_y}{\partial y}$
is the so-called divergence of $\mathbf{J}$.

An intuitive explanation why Eq.~\ref{initcond}--\ref{continuity}
imply Eq.~\ref{det} for $\mathbf{T}_\infty$ is as follows.  
Suppose a small, simply connected region $\mathcal{R}$
with area $A$ contains the point $\mathbf{r}$.
Because of the definition of the population density $\rho_0$, the
initial population contained inside $\mathcal{R}$ is approximately
equal to $A\rho_0(\mathbf{r})$.
After the boundary has drifted with the flow and reached its final
position, $\mathcal{R}$ has been mapped to a new region $\mathcal{S}$
with area $\approx A \cdot \det(\nabla\mathbf{T_\infty}(\mathbf{r}))$.
As a consequence of Eq.~\ref{lim_rho}, the population contained in
$\mathcal{S}$ is approximately $\bar\rho A \cdot
\det(\nabla\mathbf{T_\infty}(\mathbf{r}))$.
The continuity equation~\ref{continuity} guarantees that the
population inside any closed boundary is preserved while the boundary
is drifting with the velocity field~\cite{Anderson91}.
Therefore, $A\rho_0(\mathbf{r}) = \bar\rho A \cdot
\det(\nabla\mathbf{T_\infty}(\mathbf{r}))$.
After canceling the common factor $A$ on both sides and comparing
with Eq.~\ref{det}, we conclude that $\mathbf{T}_\infty$ is indeed a
density-equalizing projection.

For a rigorous proof that Eq.~\ref{det} is a consequence of
Eq.~\ref{initcond}--\ref{continuity}, we would have to impose several
demands on the continuity and integrability of $\rho$ and $\mathbf v$
so that the solution of Eq.~\ref{T_integral} is guaranteed to exist.
The technical details are beyond the scope of this article.
In general, we can safely assume that the conditions are valid in our
case.
The interested reader may consult Ambrosio et
al.\,\cite{ambrosio2008gradient} for details, especially
their Proposition 8.1.8.

\subsection*{The general solution for vortex-free flow}

Equations~\ref{initcond}--\ref{continuity} are, under mild
assumptions, sufficient to ensure that $\mathbf{T}_\infty$ is a
density-equalizing projection.
However, the equations have multiple solutions and many of them are in
practice unsuitable for producing cartograms.
In particular, solutions with vortices in the flux field $\mathbf{J}$
create severe local distortions in the vicinity of each vortex.
We therefore add one more demand to
Eq.~\ref{initcond}--\ref{continuity},
\begin{equation}
  \frac{\partial J_x}{\partial y} = \frac{\partial J_y}{\partial
    x}\ ,
  \label{novortex}
\end{equation}
which guarantees that there are no vortices~\cite{Anderson91}.

Can we construct concrete pairs of a density $\rho(x,y,t)$ and a
velocity $\mathbf{v}(x,y,t)$ that satisfy Eq.~\ref{initcond}--\ref{novortex}?
Let us assume that $\rho(x,y,t)$ is a piecewise continuous function.
At all points $(x,y)$ where $\rho(x,y,t)$ is continuous, the cosine
Fourier series of $\rho$ converges pointwise to $\rho$.
Thus, at these points we have
\begin{align}
  \rho&(x,y,t) =\nonumber\\
  &\frac1{L_xL_y} \sum_{m=0}^\infty \sum_{n=0}^\infty
  \tilde{\rho}_{mn}f_{mn}(t)\cos\left(\frac{m\pi x} {L_x}\right)
  \cos\left(\frac{n\pi y} {L_y}\right),
  \label{rhoxyt}
\end{align}
where
\begin{align*}
  \tilde\rho_{mn} = &\frac 4 {(\delta_{m0}+1)(\delta_{n0}+1)}\\
  &\times\int_0^{L_x} \int_0^{L_y} \rho(x',y',0)\cos\left(\frac{m\pi
      x'} {L_x}\right) \cos\left(\frac{n\pi y'} {L_y}\right) dx'dy'
\end{align*}
is the backward cosine Fourier transform of the initial density,
$\delta_{m0}$ is the Kronecker symbol
\[
\delta_{m0} = 
\begin{cases}
  1 & \text{if $m=0$,}\\
  0 & \text{otherwise,}
\end{cases}
\]
and $f_{mn}(t)$ is a function that must be consistent with the
constraints expressed by Eq.~\ref{initcond}--\ref{novortex}.

The functions $\cos\left(\frac{m \pi x}{L_x}\right)$ with
$m=0,1,\ldots$ are mutually orthogonal so that $f_{mn}(t)$ on the right-hand
side of Eq.~\ref{rhoxyt} is uniquely determined by $\rho(x,y,t)$
on the left-hand side.
From this observation and Eq.~\ref{initcond}, it follows that
\begin{equation}
  f_{mn}(0) = 1 \hspace{0.5cm}\text{for all $m$ and $n$.}
  \label{fmn0}
\end{equation}
Because of $\tilde\rho_{00} = \bar\rho L_x L_y$ and Eq.~\ref{lim_rho}, we must have
\begin{equation}
  \lim_{t\to\infty}f_{mn}(t) = 
  \begin{cases}
    1 & \text{if $m=n=0$,}\\
    0 & \text{otherwise.}
  \end{cases}
  \label{lim_fmn}
\end{equation}
To interpret the remaining constraints (i.e., Eq.~\ref{continuity} and
Eq.~\ref{novortex}) we must specify the boundary
conditions of the flux $\mathbf{J}$.
We assume that there is no flow through the edges of the rectangular
box $[0,L_x]\times[0,L_y]$.
Then it must be possible to express the $x$- and $y$-coordinates of the
two-dimensional function $\mathbf{J}$ in terms of the following mixed
sine and cosine Fourier transforms at all points $(x,y)$ where
$\mathbf{J}$ is continuous,
\begin{align}
  J_x(x,y,t) & = \frac1{L_xL_y}\sum_{m=1}^\infty\sum_{n=0}^\infty\tilde
  J_{x,mn}(t)\sin\left(\frac{m\pi x}{L_x}\right)\cos\left(\frac{n\pi
               y}{L_y}\right),
  \label{Jx}\\
  J_y(x,y,t) & = \frac1{L_xL_y}\sum_{m=0}^\infty\sum_{n=1}^\infty\tilde
  J_{y,mn}(t)\cos\left(\frac{m\pi x}{L_x}\right)\sin\left(\frac{n\pi
               y}{L_y}\right).
               \label{Jy}
\end{align}
We insert Eq.~\ref{rhoxyt}, \ref{Jx}, and \ref{Jy} into
Eq.~\ref{continuity}, interchange differentiation and summation, and
finally compare each term in the series on the left- and right-hand side.
The result is
\begin{equation}
  \tilde\rho_{mn} f_{mn}'(t) = -\frac{m\pi}{L_x}\tilde J_{x,mn}(t)
  -\frac{n\pi}{L_y}\tilde J_{y,mn}(t).
  \label{from_continuity}
\end{equation}
For $m=n=0$, the right-hand side is $0$ so that $f'_{00}(t) = 0$. 
From this result and Eq.~\ref{fmn0}, we can deduce that
\begin{equation}
  f_{00}(t) = 1 \hspace{0.5cm}\text{for all $t$}.
  \label{f00}
\end{equation}
Similarly, we obtain, after inserting Eq.~\ref{Jx} and \ref{Jy} into
Eq.~\ref{novortex},
\begin{equation}
  \frac n{L_y}\tilde J_{x,mn}(t) = \frac m{L_x}\tilde J_{y,mn}(t).
  \label{from_novertex}
\end{equation}
Combining Eq.~\ref{from_continuity} and Eq.~\ref{from_novertex}, we
can solve for the Fourier coefficients of the flux,
\begin{align}
  \tilde J_{x,mn}(t) & = -\frac{m L_x
                       L_y^2}{\pi(m^2L_y^2+n^2L_x^2)}\tilde\rho_{mn}f_{mn}'(t),
                       \label{Jxmn}\\ 
  \tilde J_{y,mn}(t) & =
                       -\frac{nL_x^2L_y}{\pi(m^2L_y^2+n^2L_x^2)}\tilde\rho_{mn}f_{mn}'(t).
                       \label{Jymn}
\end{align}

In summary, a flow-based density-equalizing projection is vortex-free
if and only if $f_{mn}(t)$ satisfies Eq.~\ref{fmn0}, \ref{lim_fmn},
\ref{f00} and the Fourier coefficients of the flux obey Eq.~\ref{Jxmn}
and \ref{Jymn}.

\subsection*{Equations 4--7 in the main text as a special
  density-equalizing projection with vortex-free flow}
There are many possible choices of $f_{mn}$ consistent with Eq.~\ref{fmn0},
\ref{lim_fmn} and \ref{f00}. 
The diffusion-based method of Ref.~\cite{GastnerNewman04}
corresponds to the choice 
\begin{equation}
  f_{mn, \text{diff}}(t) = \exp\left[-\left(\frac{m^2}{L_x^2} +
      \frac{n^2}{L_y^2}\right)t\right].
  \label{fmn_diff}
\end{equation}
According to Eq.~\ref{Jxmn} and \ref{Jymn}, the Fourier coefficients
of the flux are then given by
\begin{align}
  \tilde J_{x,mn,\text{diff}}(t) &= \frac m{\pi
    L_x}\tilde\rho_{mn}\exp\left[-\left(\frac{m^2}{L_x^2} + 
      \frac{n^2}{L_y^2}\right)t\right],\label{Jxmn_diff}\\
  \tilde J_{y,mn,\text{diff}}(t) &= \frac n{\pi
    L_y}\tilde\rho_{mn}\exp\left[-\left(\frac{m^2}{L_x^2} +  
      \frac{n^2}{L_y^2}\right)t\right].\label{Jymn_diff}
\end{align}

It is computationally disadvantageous that $t$ appears in the argument
of the exponential function in Eq.~\ref{fmn_diff}--\ref{Jymn_diff}.
Whenever the numerical integration of Eq.~\ref{T_integral} must
advance the time $t$ by a small increment, the Fourier 
coefficients, including the exponential function, must be computed again.
Although modern computers can evaluate the exponential function
relatively quickly, it is still slower than the four basic arithmetic
operations (i.e., addition, subtraction, multiplication, and division).
Even more time-consuming than the exponential function are the
backward Fourier transforms to $\rho(x,y,t)$ and $\mathbf{J}(x,y,t)$,
which we need in order to evaluate $\mathbf{v}$ appearing in
Eq.~\ref{T_integral}.

The alternative approach that we explore in this article is based on
the choice
\begin{equation}
  f_{mn}(t) =
  \begin{cases}
    1 & \text{if $m=n=0$,}\\
    1-t & \text{if $(m,n)\neq (0,0)$ and $0\leq t\leq 1$,}\\
    0 & \text{otherwise.}
  \end{cases}
  \label{fmn_fast}
\end{equation}
instead of Eq.~\ref{fmn_diff}. 
Performing the backward transform in Eq.~\ref{rhoxyt} shows that the
density is
\begin{equation}
  \rho(x,y,t) = 
  \begin{cases}
    (1-t)\,\rho(x,y,0) + t\bar{\rho} & \text{if $0\leq t\leq 1$},\\
    \bar\rho & \text{if $t>1$.}
  \end{cases}
  \label{rho_fast}
\end{equation}
Although the physical interpretation of the resulting flow is now less
intuitive than for the diffusion-based method, the mathematical
literature has explored solutions of the prescribed
Jacobian equation~\ref{det} based on
Eq.~\ref{rho_fast}~(\cite{dacorogna1990partial, moser1965volume}).

The Fourier coefficients of the flux follow from Eq.~\ref{Jxmn} and
\ref{Jymn},
\begin{align}
  \tilde J_{x,mn}(t) &=
                        \begin{cases}
                          \frac{m L_x L_y^2}{\pi (m^2 L_y^2 + n^2
                            L_x^2)} \tilde\rho_{mn} & \text{if $0\leq
                            t\leq 1$,}\\
                          0 & \text{otherwise.}
                        \end{cases}
  \label{Jxmn_fast}\\
  \tilde J_{y,mn}(t) &=
                       \begin{cases}
                         \frac{n L_x^2 L_y}{\pi (m^2 L_y^2 + n^2
                           L_x^2)} \tilde\rho_{mn} & \text{if $0\leq
                           t\leq 1$,}\\
                         0 & \text{otherwise.}
                       \end{cases}
                             \label{Jymn_fast}
\end{align}
Upon inserting Eq.~\ref{Jxmn_fast} and \ref{Jymn_fast} into
Eq.~\ref{Jx} and \ref{Jy}, we obtain the flux
\begin{align}
  J_x(x,y,t) = -\frac{L_y}\pi &\sum_{m=1}^\infty \sum_{n=0}^\infty
                                \bigg[\frac m{m^2L_y^2+n^2L_x^2}
                                \tilde\rho_{mn}
                                \nonumber\\
                              &\times\sin\left(\frac{m\pi
                                x}{L_x}\right) \cos\left(\frac{n\pi y}{L_y}\right)\bigg]\ ,
                                \label{Jx_fast}\\
  J_y(x,y,t) = -\frac{L_x}\pi &\sum_{m=0}^\infty \sum_{n=1}^\infty
                                \bigg[\frac n{m^2L_y^2+n^2L_x^2}
                                \tilde\rho_{mn}
                                \nonumber\\
                              &\times\cos\left(\frac{m\pi
                                x}{L_x}\right) \sin\left(\frac{n\pi y}{L_y}\right)\bigg]
                                \label{Jy_fast}
\end{align}
for $0\leq t\leq 1$. For $t>1$, we simply get $J_x=J_y=0$.
When we divide $J_x$ and $J_y$ by the density $\rho$ in
Eq.~\ref{rho_fast}, we obtain equations 5 and 6 in the main text.

There are multiple advantages when choosing Eq.~\ref{fmn_fast} instead
of Eq.~\ref{fmn_diff}. 
\begin{itemize}
\item As we have just derived from Eq.~\ref{fmn_fast}, the flux is
  zero after $t=1$.
  It follows that $\mathbf  T_1=\mathbf T_\infty$.
  Hence, there is no need to take the limit $t\to\infty$ when we
  perform the integral in Eq.~\ref{T_integral}. 
  In practice, we no longer need to apply heuristics to test whether the
  integrand at time $t$ is small enough to terminate the numerical
  integration.
  Instead, we integrate until the fixed upper integration limit $t=1$,
  which is easier to implement.
\item Unlike in the diffusion-based method, we can calculate the
  density $\rho(x,y,t)$ in Eq.~\ref{rho_fast} without Fourier transforms.
\item
  In the diffusion-based method, the Fourier coefficients of the flux 
  (Eq.~\ref{Jxmn_diff} and \ref{Jymn_diff}) are time-dependent.
  Therefore, at every new time step during the numerical integration
  of Eq.~\ref{T_integral}, we must carry out a new backward Fourier
  transform.
  By contrast, the right-hand sides of Eq.~\ref{Jx_fast} and
  \ref{Jy_fast} do not depend on $t$. 
  It suffices to perform the summations once at the start of the
  algorithm, most efficiently with the fast Fourier transform
  technique~\cite{FrigoJohnson05}.
  If we store the result in memory, we do not need any more Fourier
  transforms at all during the integration.
\item After computing the sums in Eq.~\ref{Jx_fast} and \ref{Jy_fast}
  at the beginning of the code, we only need addition,
  subtraction, multiplication, and division.
  In particular, we never need to evaluate the exponential function
  that appears in Eq.~\ref{fmn_diff} of the diffusion-based method.
\end{itemize}
The overall effect is remarkably fast computer code. 
For typical runs, we find that a serial implementation of the
algorithm based on Eq.~\ref{fmn_fast} only takes around $18\%$ of the
time needed for the diffusion-based method.
By parallelizing the integrator, it is possible to speed up the code
even further.
With a 12-core processor, we were able to reduce the time needed by
the new algorithm to only around $3\%$ of the run-time for the
diffusion-based code.

\section{Tissot ellipses and angular-distortion metrics}

In cartography, the Tissot indicatrix is a visual and numerical concept to analyze the distortions generated by a map projection. Introduced by Nicolas Auguste Tissot in the nineteenth century, the Tissot indicatrix has become an important tool, especially when characterizing projections of the Earth's (nearly) spherical surface onto a two-dimensional plane.
The framework of our article is different: we are transforming a two-dimensional map (the cartogram input) to another two-dimensional map (the cartogram).
Still, we can use Tissot indicatrices to measure the magnitude of the distortions produced by different cartogram algorithms.

\subsection*{Tissot ellipses}
\begin{figure*}
\includegraphics[width=\linewidth]{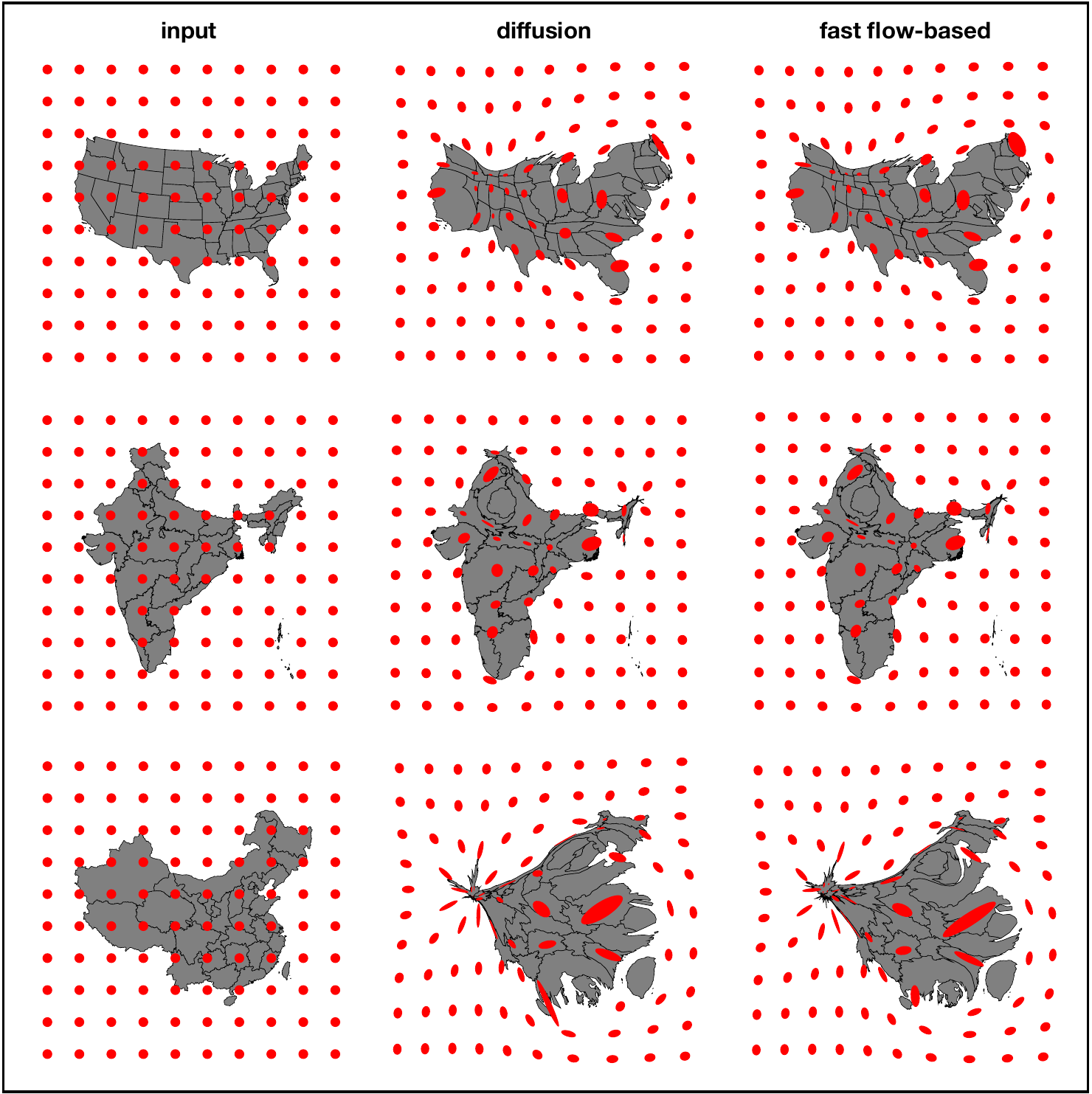}
\caption{
Tissot indicatrices obtained for the diffusion-based algorithm (middle column) and the fast flow-based algorithm proposed in the main text (right column). 
The unprojected circles are displayed in the left column. 
The cartograms for the USA (top row), India (middle row), mainland China and Taiwan (bottom row) are based on the same data as Fig.\,3, 4, and 5 of the main text.
}
\label{tissot}
\end{figure*}

Consider an infinitesimally small circle centered at $(x, y)$ on the input map.
Locally, a smooth map projection $\mathbf T$ is approximately equal to the affine transformation 
\begin{equation}
\mathbf T(x+\delta_x, y+\delta_y) \approx \mathbf T(x, y) + \nabla \mathbf T(x, y)
\begin{pmatrix}
\delta_x\\
\delta_y
\end{pmatrix}\ ,
\label{eq:Taffine}
\end{equation}
so long as $\delta_x$ and $\delta_y$ are sufficiently small.
Here $\nabla \mathbf T$ is the Jacobian matrix.
It can be shown that any affine transformation applied to a circle results in an ellipse. 
The Tissot indicatrix of $(x,y)$ under the projection $\mathbf T$ is the ellipse generated by the affine transformation on the right-hand side of Eq.~\ref{eq:Taffine} when applied to the infinitesimal circle at $(x,y)$.
In the left-hand column of Fig.~\ref{tissot}, we show several circles placed at regularly spaced locations on the input maps of our benchmarking examples (USA by electors, India and China by GDP).
We use a finite radius to make the circles visible.
In the middle and right-hand columns, we show the corresponding Tissot indicatrices centered at locations $\T(x,y)$ on diffusion and fast-flow based cartograms, respectively.

\subsection{Angular-distortion metrics}
It is desirable for a density-equalizing projection $\T$ to preserve shapes as much as possible so that each area is easily recognizable by the reader of the cartogram. 
One way to interpret shape preservation is to demand that angles remain locally unchanged by the transformation $\T$. 
This property is referred to as \textit{conformality}. 
Equivalently, a conformal transformation must satisfy both Cauchy-Riemann equations  
\[
    \frac{\partial T_x}{\partial x} = \frac{\partial T_y}{\partial
      y}, \hspace{0.5cm} \frac{\partial T_x}{\partial y} = -\frac{\partial
      T_y}{\partial x}\ .
\]
Together with the prescribed Jacobian equation (1) in the main text, a conformal density-equalizing projection would have to satisfy three equations.
In general, two functions $T_x$ and $T_y$ cannot satisfy three independent constraints so that a perfectly conformal density-equalizing solution is infeasible~\cite{GuseinTikunov93}.
Yet, a visually pleasing cartogram should deviate from conformality as little as possible. 
Although our present paper focuses on building a fast cartogram algorithm rather than achieving small conformality error, we compute several conformality metrics to verify that our proposed algorithm produces cartograms that are in this respect as good as the state-of-the-art diffusion algorithm.

Angular distortion metrics can be derived from the properties of Tissot ellipses.
Consider the Tissot ellipse that is the image of the unit-radius circle centered at $(x,y)$ after applying the affine transformation of Eq.~\ref{eq:Taffine}.
We denote the length of the ellipse's semi-major and semi-minor axis by $a(x,y)$ and $b(x,y)$, respectively. 
In the case of a conformal projection $\T$, we would have $a(x,y) = b(x,y)$. That is, the Tissot ellipse would be a circle whose radius can be smaller or bigger than $1$. 
Hence, we can define a measure of the angle distortion at $(x,y)$ by
\begin{equation}
	e(x,y) = \ln\left(\frac{a(x,y)}{b(x,y)}\right),
    \label{eq:e}
\end{equation}
as described for example in Ref.~\cite{Papadopoulos17}. 
We choose the average
\[
e_a = \frac{1}{|\Omega|}\int_\Omega \ln\left(\frac{a(x,y)}{b(x,y)}\right)dx\,dy,
\]
and the largest value
\[
e_\infty = \underset{\x \in \Omega}{\sup} \ln\left(\frac{a(x,y)}{b(x,y)}\right)
\]
as two global measures for the distortion error, where $\Omega$ is the total area of the cartogram.
Here we choose $\Omega$ as the $L_x\times L_y$ bounding rectangle described in the section ``Benchmarking the algorithm with data for the USA, India, and China'' in the main text.
Another angular-distortion metric can be computed from the local maximum angular-value change (see derivation in \cite{snyder1987map}),
\begin{equation}
	\tilde{e}(x,y) = 2 \arcsin \left( \frac{a(x,y)-b(x,y)}{a(x,y)+b(x,y)}\right),
    \label{eq:etilde}
\end{equation}
which also provides two global angular-distortion metrics $\tilde{e}_a$ and $\tilde{e}_\infty$.
We display these errors for both the diffusion-based and our new proposed algorithm in Table 1 of the main text.

\section{Polygon-level distortions}

The metrics $e$ and $\tilde e$ defined in Eq.~\ref{eq:e} and~\ref{eq:etilde} are local: they can be computed only for {\it ``all-coordinates''} cartograms for which we know the transformation $\T$ at every location $(x,y)$. 
In order to obtain metrics that are well-defined for general cartograms, one has to measure distortions at the level of polygons instead of all coordinates. 
Such metrics have been introduced in several previous articles~\cite{Alam_etal15, keim2003, heilmann2004recmap}. 
According to some metrics, the fast flow-based algorithm defined in the main text and several other contiguous methods, including the diffusion cartogram, are already optimal.
For example, both the diffusion and fast flow-based algorithm succeed in rescaling the regions to their objective areas and preserve the adjacency between polygons.
Three metrics that meaningfully compare the diffusion and fast-flow based algorithm are (1) the average aspect ratio $\alpha$, (2) the total Hamming distance $\delta$, and (3) the relative position error $\theta$.
We describe them below and display the results for each cartogram in Table 1 of the main text.

\subsection*{Average aspect ratio}
Cartograms in which polygons become thin and elongated are difficult to read.
It is also difficult to place labels inside such polygons.
The aspect ratio of a polygon quantifies how stretched it appears.
For the $i$-th polygon on the cartogram, we define $l_i(\phi)$ and $s_i(\phi)$ as the longer and shorter side length, respectively, of the bounding rectangle whose edges are at angles $\phi$ and $\phi+90^\circ$ with respect to the $x$-axis.

We define $\phi_{\min, i}$ as the angle at which the bounding rectangle for the $i$-th polygon has the minimum area,
$$\phi_{\min, i} = \underset{\phi\in[0^\circ,90^\circ]}{\arg \min} [l_i(\phi)\cdot s_i(\phi)].$$
The aspect ratio of the $i$-th polygon is the ratio of the larger to the smaller side length of the bounding rectangle of minimum area for that polygon,
\[
\alpha_i = \frac{l_i(\phi_{\min, i})} {s_i(\phi_{\min, i})}.
\]
If there are $p$ polygons on the cartogram, we define $\alpha$ as the mean aspect ratio,
\[
\alpha = \frac 1 p \sum_{i=1}^p \alpha_i\ .
\]

\subsection*{Total Hamming distance} 
The Hamming distance $h$ measures the difference in the shapes of two polygons. 
It is computed by superimposing one polygon on top of another and measuring the fraction of area that lies in only one, but not both polygons,
\[
h = \frac{\text{area in exactly one polygon}}{\text{sum of areas of individual polygons}}\ .
\]
In our application, one of the polygons is from the input map, the other is the corresponding polygon from the cartogram.
We rescale the cartogram polygon so that it has the same area as the polygon before the cartogram projection. 
Otherwise we would unfairly penalize cartograms that correctly changed the polygon areas to their objective values.
To make the measure translation invariant, we define $\delta_i$ as the minimum Hamming distance of all possible translations of the rescaled $i$-th cartogram polygon with respect to the $i$-th unprojected polygon. 
The total Hamming distance $\delta$ is obtained by summing the Hamming distances of all polygons.

\subsection*{Relative position error}
We can quantify changes in the relative position of two polygons $i$ and $j$ between input map and cartogram by measuring the angle $\phi_{ij}$ between the lines connecting the centroids before and after the projection.
If $\mathbf c_i$, $\mathbf c_j$ are the centroids on the input map and $\mathbf d_i$, $\mathbf d_j$ on the cartogram, then 
\[
\phi_{ij} = \arccos\left(\frac{(\mathbf c_i - \mathbf c_j)\cdot(\mathbf d_i - \mathbf d_j)}{|\mathbf c_i - \mathbf c_j|\cdot|\mathbf d_i - \mathbf d_j|}\right)\ .
\]
We define the relative position error $\theta$ as the average of $\phi_{ij}$ over all possible pairs of polygons.
We also divide by $\pi$,
\[
\theta = \frac 2{p(p-1)\pi} \sum_{i=1}^{p-1} \sum_{j=i+1}^p \phi_{ij},
\]
so that $\theta\in[0,1]$~\cite{heilmann2004recmap}.

\end{document}